\documentclass{article}

\usepackage[utf8]{inputenc}
\usepackage{amsmath}
\usepackage{amssymb}
\usepackage{amsthm}
\usepackage{authblk}
\usepackage{comment}
\usepackage{float}
\usepackage{graphicx}
\usepackage{hyperref}
\usepackage{listing}
\usepackage{listings}
\usepackage{mathtools}
\usepackage{multirow}
\usepackage{tabularx}
\usepackage{tabulary}
\usepackage{url}
\usepackage{xcolor}

\definecolor{mygreen}{RGB}{28,172,0} 
\definecolor{mycyan}{RGB}{0,255,255}
\definecolor{mylilas}{RGB}{170,55,241}
\definecolor{myazure}{RGB}{0,175,225}
\definecolor{myorange}{RGB}{175,125,0}
\definecolor{chatbackcolor}{rgb}{0.95,0.95,0.92}

\lstnewenvironment{mm}[1][]{%
  \lstset{
  language=Matlab,
  backgroundcolor=\color{chatbackcolor},
  showstringspaces=false,
  xleftmargin = 0cm,
  xrightmargin = 0cm,
  extendedchars=true,
  captionpos=b,
  basicstyle=\small\rmfamily,
  tabsize=2,
  keywordstyle=\textbf,
  stringstyle=\textit,
  numbers=none,
  numberstyle=\small,
  breakautoindent  = true,
  breakindent      = 2em,
  breaklines       = true,
  postbreak        = ,
  frame=single,
  keywordstyle=\color{blue},
  numbers=left,%
    numberstyle={\tiny \color{black}},
    numbersep=9pt, 
    emph=[1]{function,for,end,break},
    emphstyle=[1]\color{red}, 
  literate={ö}{{\"o}}1
           {ä}{{\"a}}1
           {ü}{{\"u}}1, #1
}}{}

\urldef{\mailAM}\path|alexmos@kma.zcu.cz |
\urldef{\mailTR}\path|talal.rahman@hvl.no |
\urldef{\mailJV}\path|jvaldman@jcu.cz |
\urldef{\mailJEV}\path|jon.e.vatne@bi.no |

\theoremstyle{definition}
\newtheorem{remark}{Remark}
\newtheorem{example}{Example}

\newcommand{\R}{\mathbb R}

\renewcommand{\d}{d}
\renewcommand{\dim}{dim}

\newcommand{\nt}{|\mathcal{T}|}
\newcommand{\nn}{|\mathcal{N}|}

\newcommand{\nfb}{|\mathcal{F}_b|}
\newcommand{\nip}{n_{ip}}

\newcommand{\Hom}{\operatorname{Hom}}
\newcommand{\ten}[1]{\mathcal{#1}}
\newcommand{\dual}{{}^\vee}
\newcommand{\eval}{\operatorname{eval}}

\newcommand{\Id}{\operatorname{Id}}
\newcommand{\RR}{\mathbb{R}}
\newcommand{\CC}{\mathbb{C}}

\newcommand{\dxdydz}{\mathrm{d}\boldsymbol{\mathrm{x}}}

\newcommand{\dS}{\, \mathrm{d}S}
\newcommand{\x}{\boldsymbol{\mathrm{x}}}
\newcommand{\ksi}{\boldsymbol{\mathrm{\xi}}}

\newcommand{\uL}{u}

\newcommand{\cK}{c_K}
\newcommand{\cM}{c_M}
\newcommand{\cKx}{\cK(\x)}
\newcommand{\cMx}{\cM(\x)}

\newcommand{\vx}{v(\x)}
\newcommand{\IK}{I_K}
\newcommand{\IM}{I_M}
\newcommand{\eK}{e_K}
\newcommand{\eM}{e_M}
\newcommand{\ux}{u(\x)}
\newcommand{\uDx}{u_D(\x)}
\newcommand{\fx}{f(\x)}
\newcommand{\Ja}{J_1}
\newcommand{\Jb}{J_2}
\newcommand{\Jc}{J_3}
\newcommand{\Jak}{J_{1,k}}
\newcommand{\Jbk}{J_{2,k}}
\newcommand{\Jck}{J_{3,k}}

\newcommand{\unum}{u_h}
\newcommand{\unumx}{\unum(\x)}
\newcommand{\uvec}{\tilde{u}}
\newcommand{\fvec}{\tilde{f}}
\newcommand{\Kk}{K_k}
\newcommand{\Mk}{M_k}
\newcommand{\bk}{b_k}
\newcommand{\uveck}{\uvec_k}
\newcommand{\matlab}{MATLAB }

\newcommand{\textbfn}[1]{'\textbf{#1}'}
\newcommand{\textn}[1]{'{#1}'}

\newcommand{\coords}{\textbfn{coords3D}}
\newcommand{\vecs}{\textbfn{vectors3D}}
\newcommand{\facesB}{\textbfn{facesB}}
\newcommand{\areasB}{\textbfn{areasB}}

\title{On a vectorized basic linear algebra package for prototyping codes in \matlab}

\author[1,5]{Alexej Moskovka}
\author[2]{Talal Rahman}
\author[3,5]{Jan Valdman}
\author[4]{Jon Eivind Vatne}
\affil[1]{Department of Mathematics, Faculty of Applied Sciences,
University of West Bohemia, Technick\' a 8, 30100 Pilsen, Czechia,
\mailAM}

\affil[2]{Faculty of Engineering and Science, Western Norway
University of Applied Sciences, Inndalsveien 28, 5063 Bergen, Norway,
\mailTR}

\affil[3]{Department of Computer Science, Faculty of Science, University of South Bohemia, 
Brani\v sovsk\' a 31, 37005~\v{C}.~Bud\v{e}jovice, Czechia,
\mailJV}

\affil[4]{Department of Economics, BI Norwegian
Business School, Kong Christian Frederiks plass 5, 5006 Bergen, Norway,
\mailJEV}

\affil[5]{The Czech Academy of Sciences, Institute of Information Theory and Automation, Pod Vod\'{a}renskou v\v{e}\v{z}\'{\i}~4, 18208~Prague, Czechia.}

\date{}

\begin{document}

\maketitle

\begin{abstract}

When writing a high-performance code for numerical computation in a scripting language like \matlab, it is crucial to have the operations in a large for-loop vectorized. If not, the code becomes too slow to be of any use, even for a moderately large problem. However, in the process of vectorizing, it often happens that the code loses its original structure and becomes less readable. This is particularly true in the case of a finite element implementation, even though finite element methods are inherently structured. A basic remedy to this is the separation of the vectorization part from the mathematics part in the code, which is easily achieved through building the code on top of the basic linear algebra subprograms that are already vectorized codes, an idea which has been used in a series of papers over the last fifteen years, developing codes that are fast and still structured and readable.  We discuss the vectorized basic linear algebra package, and introduce
a formalism using multi-linear algebra to explain and define formally the functions in the package, as well as \matlab's pagetime functions.       
We provide examples from computations of varying complexity, including the computation of normal vectors, volumes, and finite element methods. Benchmarking shows that we also get fast computations. Using the library, we can write codes that closely follow our mathematical thinking, making it easier to write, follow, reuse, and extend the code. 
\end{abstract}

\tableofcontents

\section{Introduction}

\matlab \cite{MATLAB} is a popular computing platform with a library of built-in functions and toolboxes provided by Mathworks Inc. (\url{ https://mathworks.com/}), to solve scientific and engineering problems in both academia and industry. It is a scripting language that can be used to write structured and readable or understandable code. However, because the language is interpreted and not compiled, \matlab codes containing for loops become extremely slow compared to compiled languages such as C, C++, FORTRAN, etc. 
It is particularly evident in a finite element calculation, since a finite element implementation is heavily based on loops over its nodes, edges, and elements; cf. \cite{AlbertyCarstensenFunken1999}, making any large-scale simulation with finite elements practically useless even on a supercomputer. \matlab provides functionalities that allow basic arithmetic operations in a loop to be executed in a precompiled fashion, also known as the vectorization or array operation \cite{MATLAB}. 
During the last 15 years, a number of finite element codes have been developed in \matlab using vectorization to speed up calculations; cf. \cite{Cuve2016, Koko2007}. 
Vectorized FE codes show a tremendous improvement in the performance of their time to compute compared to their non-vectorized version. However, in a straightforward vectorized version, as the mathematics becomes interleaved with the array operations, the code quickly loses its structure and readability, making it hard to use it in a class room or in real applications to further develop or extend. To retain readability and structure, one needs to think in a whole new way, one of which is the separation of vectorization from the mathematics of the problem, an idea which was first introduced for the implementation of finite elements in the numerical simulation of the Electro-Rheological Fluid \cite{Lit2005, Lit2007}, and documented in \cite{Rahman2003}. In this idea, the vectorization was done by extending the element-wise operations into matrix-wise or page-wise operations.     

In Rahman and Valdman \cite{RahmanValdman2013}, the authors used the idea to further develop an efficient and flexible assembly procedure for the FEM stiffness and mass matrices for nodal elements in 2D and 3D. This resulted in a faster and more scalable algorithm that led to a significant speed-up of the original codes (cf. \cite{AlbertyCarstensenFunken1999}). The same idea of vectorization was also used in the assembly of edge elements by Anjam and Valdman in the \cite{AnjamValdman2015} and $C^1$ elements by Valdman \cite{Valdman2020}. The ability to simultaneously create FEM matrices for problems formulated in Sobolev spaces $H^1, H(\operatorname{div}), H(\operatorname{curl}), H^2$ resulted in additional \matlab related computations, including a posteriori functional estimates
\cite{AnjamValdman2015, BozorgniaValdman2017}
generalized eigenvalue problems \cite{PaulyValdman2020} and models in the continuum mechanics of solids \cite{FriedrichKruzikValdma2021,KroemerValdman2019}. 
The idea of original vectorization offers more features that were only partially explored. An example is \cite{MarcinkowskiValdman2020} that attempted simple iterative solvers for solutions of the Laplace equation, completely avoiding the setup of stiffness and mass matrices. 

The aim is to develop a mathematical framework for the library, the vectorized basic linear algebra package, to be able to formally define the functions in the library, as well understand their constructions so as to be able to write better code.
We include some background material from the linear (and multilinear) algebra, where the selection of what to include is based on the library functions, \matlab's page wise functions, and the applications we have in mind.
In particular, tensors play a central role.
The library contains a collection of page-wise vectorized versions of functions which are Basic Linear Algebra Subprograms (abbreviated as BLAS). The library enables us to separate the vectorization from the mathematical method or algorithm, e.g. the FE or the geometry algorithm, and hence make the final code readable and reusable.
While we were preparing examples for this paper, working with the linear algebra at a higher level (i.e. above the BLAS level), the vectorization being implicit, we can make a clearer and a much more structured code. 
We provide several examples in detail.
From geometry, these include computations of normal vectors and of volume.
From FEM, our main motivation, we provide examples for different kinds of elements.
In all cases, the code is strongly linked to a linear algebra formulation. 
In most cases, we have left this implicit, but in some cases, as a guide to the reader, we have made this explicit; see, e.g., Remarks \ref{rem:coords3D} and \ref{rem:phider}.


All \matlab codes used in this article can be downloaded at \url{https://www.mathworks.com/matlabcentral/fileexchange/130824}.
We note that the library itself is written in \matlab, so it is possible to read and modify if one desires.

\section{Background from linear algebra}
The present goal is to introduce the necessary tools of linear algebra to understand Matlab vectorization. We will make both natural and coordinate-dependent constructions. When we describe the main functions of the library in Section \ref{subsec:lib}, we refer to the linear algebra construction underlying the functions. Later, we will give examples of how to see the connection between on the one hand the code using the library, and on the other hand the linear algebra; see, for instance, Remarks \ref{rem:coords3D} and \ref{rem:phider}. We make this connection explicit only in select cases, though this reasoning permeates all our constructions.

Let $U,V,W,...$ be real vector spaces which we usually assume to be finite dimensional.
Let $u\in U,\,v\in V,\dots$ denote arbitrary vectors in the given vector spaces.
As a general reference for linear algebra, we suggest \cite{Simo}.
See e.g. \cite{MacL} for a general exposition of tensor products, their connections with homomorphism spaces and much more.
Some of our constructions are usually formulated in a more general setting, as in \cite{MacL}, but we have freely made the simplifications that are possible since we work only with vector spaces.

\subsection{Homomorphism spaces and tensor products}
The linear space of the transformations from $U$ to $V$ is

\begin{equation}
\Hom(U,V) =\{\text{linear transformations }U\rightarrow V\}.
\end{equation}
For any vector space $U$, we have an isomorphism
\begin{equation}
\label{ChooseElement}
\Hom(\RR, U)\simeq U, \phi \mapsto \phi(1).
\end{equation}
The dual space of $U$ is
\begin{equation}
\label{defDual}
U\dual = \Hom(U,\RR).
\end{equation}
There is a natural map
\begin{equation}
\label{doubleDual}
U\rightarrow U\dual \dual,\,u \mapsto \eval_u, \eval_u(\phi)=\phi(u).
\end{equation}
This map is an isomorphism if and only if $U$ is finite-dimensional.

The adjoint linear transformation to $\phi\in \Hom(U,V)$ is the map $\phi^T\in \Hom(V\dual, U\dual)$ given by
\begin{equation}
\label{transpose}
\phi^T(f)(u) =f(\phi(u)).
\end{equation}
In words: $\phi^T$ sends a map $f$ from $V$ to $\RR$ to its precomposition with $\phi$, thus giving a map from $U$ to $\RR$.\\
When $U=V$ there is a special {\em identity map} $\Id_U\in \Hom(U,U)$ defined by $\Id_U(u)=u$.

\subsubsection*{Basis}
As we aim towards numerics, it is sensible to express vectors in terms of numbers.
This is typically achieved by choosing a basis for each vector space we consider.
A choice of basis determines several isomorphisms (again, keep in mind that our vector spaces are finite dimensional).
First some notation. If the vectors $f_1,\dots,f_n$ form a basis for the vector space $U$, we write
\begin{equation}
    U=<f_1,\dots,f_n>.
\end{equation}
Let the standard basis for $\mathbb{R}^n$ be $e_1,\dots,e_n$.
Then a choice of basis for $U$ determines an isomorphism
\begin{equation}
    U\rightarrow \mathbb{R}^n,\,\,\sum a_if_i\mapsto \sum a_ie_i.
\end{equation}
It also determines a basis for the dual space $U^\vee$, namely
\begin{equation}
    U^\vee=<f^1,\dots,f^n> \quad\text{where}\quad f^j\left(\sum a_if_i\right)= a_j,
\end{equation}
and an isomorphism
\begin{equation}
    U\rightarrow U^\vee,\,\,\sum a_if_i\rightarrow \sum a_if^i.
\end{equation}

Note that if we let $U$ be an infinite dimensional vector space, $U$ and $U^\vee$ are not isomorphic.

The space of transformations from $U=<f_1,\dots,f_n>$ to $V=<g_1\dots,g_m>$ is isomorphic with the space of matrices of size $m\times n$.
If $T$ is a linear transformation, it is identified with the matrix
\begin{equation}
\label{eq:transformationMatrix}
 \big(T(f_1)|T(f_2)|\dots |T(f_n)\big).
\end{equation}

Here the columns are the images of the basis vectors in $U$ expressed in the basis for $V$.
If we express $\sum a_if_i$ as a column vector $(a_i)$, the image of this vector under $T$ is given by the matrix product $ \big(T(f_1)|T(f_2)|\dots |T(f_n)\big)(a_i)$.
Also, the adjoint from \eqref{transpose} corresponds to matrix transposition.

\subsubsection*{Tensor products}
The tensor product of two vector spaces $U$ and $V$ is 
\begin{equation}
U\otimes V=\{\text{linear combinations of }u\otimes v\}/ \text{bilinear relations}.
\end{equation}
We have isomorphisms
\begin{equation}
\label{swap}
U\otimes V \simeq V\otimes U,u\otimes v\mapsto v\otimes u
\end{equation}
and
\begin{equation}
\label{squeeze}
\RR\otimes V \simeq V,r\otimes v \mapsto rv.
\end{equation}

Choosing bases $U=<f_1,\dots,f_n>$ and $V=<g_1\dots,g_m>$ determines the basis
\begin{equation}
    U\otimes V = <f_i\otimes g_j>,\,i\in \{1,\dots, n\}, j\in\{1,\dots, m\}.
\end{equation}
As a consequence, $U\otimes V$ is isomorphic to the space of matrices of size $n\times m$, where 
\begin{equation}
\label{eq:tensorMatrix}
    \sum_{i,j} a_{ij}f_i\otimes g_j\mapsto \left(a_{ij}\right)_{ij}
\end{equation}

The map \eqref{swap} corresponds to matrix transposition, and the map \eqref{squeeze} to considering a matrix of size $1\times m$ as a vector of size $m$, as in the \matlab command \verb+squeeze+.

\subsubsection*{Connections between tensors and homomorphisms}

Composition of linear transformations is a bilinear operation, so that there is a natural map

\begin{equation}
\label{composition}
\Hom(U,V)\otimes \Hom (V,W)\rightarrow \Hom(U,W),\,\phi\otimes\psi \mapsto\psi\circ \phi.
\end{equation}
Given two maps $\phi_i\in \Hom(U_i,V_i),\,i=1,2$, we can form the tensor product map $\phi_1\otimes \phi_2\in \Hom(U_1\otimes U_2,V_1\otimes V_2)$ by defining
\begin{equation*}
(\phi_1\otimes \phi_2)(u_1\otimes u_2) = \phi(u_1)\otimes \phi(u_2).
\end{equation*}
A case we will use repeatedly is when one of the two maps is the identity map.

An important connection between tensors and homomorphisms is adjunction:
\begin{equation}
\label{adjunction}
\Hom(U\otimes V, W)\simeq \Hom(U,\Hom(V,W)),\,\phi\mapsto \psi, \psi(u)(v) = \phi(u\otimes v).
\end{equation}
Here $\phi$ is a function of two variables, whereas $\psi$ is a function of one variable,  whose value $\psi(u)$ is again a function of one variable.
In the special case $W=\RR$, combining \eqref{defDual} and \eqref{adjunction} gives
\begin{equation}
\label{eq:dualAdjunction}
(U\otimes V)\dual \simeq \Hom(U,V\dual).
\end{equation}
The following map gives an isomorphism (again, remember that we have assumed finite dimension):
\begin{equation}
\label{tensorHomInterchange}
U\otimes V\dual \simeq \Hom(V,U),\,u\otimes\phi \mapsto \psi,\psi(v) = \phi(v)u.
\end{equation}
In particular, since any finite dimensional vector space is isomorphic to a dual space, this allows any question about tensor products to be formulated using homomorphisms, and vice versa.
The matrix representations in \eqref{eq:transformationMatrix} and \eqref{eq:tensorMatrix} are compatible under these reformulations.
In the special case $V=U$, we find that
\begin{equation}
U\otimes U\dual \simeq \Hom(U,U).
\end{equation}
The element in $U\otimes U\dual$ that corresponds to $\Id_U$ under this isomorphism is called the {\em trace element}.
Its matrix representation is the identity matrix.

\subsubsection*{Bilinear forms}

A linear map from $U\otimes U$ to $\RR$ is called a bilinear form. It is customary to write this as $<-,->\in \Hom(U\otimes U,\RR)$, especially if it is non-degenerate. 
It is seldom problematic that this notation looks similar to a vector space with a basis of two vectors.
Any bilinear form induces a map $U\rightarrow U\dual$ under the map we get from \eqref{eq:dualAdjunction}:
\begin{equation}
\label{eq:bilinearGivesMap}
(U\otimes U)^\vee\simeq \Hom (U,U\dual).
\end{equation}
Concretely, given $<-,->$, the element $u\in U$ is mapped to the function $<u,->\in U\dual$.
This gives an isomorphism $U\simeq U^\vee$ precisely when the bilinear form is non-degenerate.
Even though there is much more to be said about this, we will not need anything except simple special cases.
In particular, if the bilinear form is the standard inner product on $\mathbb{R}^n$, and the standard basis is used to identify $\mathbb{R}^n$ with its dual space, the isomorphism we get from \eqref{eq:bilinearGivesMap} is the identity map.

\subsubsection*{Diagonals}
For any vector space $U$, there is a map $\Delta\in \Hom(U, U\otimes U)$ given by $u\mapsto u\otimes u$.
The image of this map can be thought of as a diagonal in $U\otimes U$.
If we identify $U\otimes U$ with the space of $n\times n$ matrices, the image is exactly the set of diagonal matrices.
With a choice of basis $U=<f_1,\dots,f_n>$, we can also consider this map
\begin{equation}
\label{eq:DiagonalVector}
\pi\in \Hom(U\otimes U,U),f_i\otimes f_i \mapsto f_i, f_i\otimes f_j\mapsto 0 \text{ if } i\neq j
\end{equation}
In the matrix representation of a tensor product, $\pi$ picks out the diagonal as a vector.


\subsubsection*{Higher tensors}
Since tensor products and homomorphism spaces of vector spaces again are vector spaces, everything we have talked about can be iterated to cover more than two factors.
For the applications we have in mind, we will usually need three factors, but sometimes more. We will keep in mind the straight-forward generalization, but focus on triple tensor products like
\begin{equation}
\ten{A} = U\otimes V\otimes W.
\end{equation}
By choosing bases for each factor, we can get a basis for the triple (or higher) tensor product by triples (or higher) of basis elements.
This allows for representing an element in $\ten{A}$ by its three-dimensional array of coefficients.
In the special case where one of the spaces is one-dimensional, we can use Equation~\eqref{squeeze} to remove the one-dimensional space, and thus reduce the dimension of the array. In \matlab, this is handled by the \verb+squeeze+-command.\\ 

In many cases, we will consider operations that are composed from simpler maps only acting on two of the three factors, with the identity map used in the third factor.
We will also freely move between tensor formulations and homomorphism space formulations, and mix these in spaces like $U\otimes \Hom(V,W)$, which is also represented by a three-dimensional array.

\subsection{Indexing and page-wise operations}
We will consider a number of maps, used in the implementation, in their abstract setting.
A common feature in many of these examples is that we want to consider one tensor factor $W$ as a method for indexing, and then perform ordinary matrix operations on the other factors.
When using array representations, we can think of this as performing operations on layers of the array.
Not much is lost if the reader wishes to think about $W=\RR^n$, but we will sometimes need a different interpretation of $W$.
\subsubsection*{Copying a vector space}
Consider the map $1_n\in \Hom(\RR,\RR^n)$ that sends $1$ to the vector $(1,1,\dots,1)^T$.
For any vector space $U$, we get a map:
\begin{equation}
\label{eq:copy}
\operatorname{copy}=\Id\otimes 1_n: U\simeq U\otimes \RR \rightarrow U\otimes \RR^n.
\end{equation}
We can think of this as giving $n$ copies of an element in $U$.
E.g. if $U\simeq \RR^m$ consists of column vectors, each column vector is sent to the matrix all of whose columns are equal to the original vector.\\
Similarly, we can extract the part corresponding to a given index by tensoring with the map $\RR^n\rightarrow \RR$ that is given by the dual basis element in $\RR^\vee$.
For instance, the last part can be extracted by

\begin{equation}
\label{eq:last}
\operatorname{last}=\Id\otimes (0\, 0 \,\dots \,0\, 1): U\otimes \RR^n \rightarrow U\otimes \RR\simeq U.
\end{equation}

\subsubsection*{Page-wise matrix multiplication}

We will consider vector spaces $\ten{A},\ten{B}$ where
\begin{equation}
\ten{A} = \Hom(V,U)\otimes W_1,\quad \ten{B} =\Hom(U,X)\otimes W_2.
\end{equation}
Using \eqref{swap} and \eqref{composition} we get a map
\begin{equation}
\label{eq:pagewisemultGeneral}
\ten{A}\otimes\ten{B}\rightarrow \Hom(U,X)\otimes W_1\otimes W_2
\end{equation}
If the case where $W=W_1=W_2$ and have a choice of basis for $W$, we can then apply the map $\pi$ defined in \ref{eq:DiagonalVector} to get a map
\begin{equation}
\label{layeredMultiplication}
 \Hom(V,U)\otimes W\otimes \Hom(U,X)\otimes W\rightarrow \Hom(V,X)\otimes W.
\end{equation}

We will think about this as a page-wise matrix multiplication: For a fixed basis element of $W$, the corresponding map is just ordinary matrix multiplication, which is then extended linearly.
This interpretation is exactly what happens when we represent $\ten{A}$ and $\ten{B}$ as three-dimensional arrays in \matlab and use the command \verb+pagemtimes+.

\subsubsection*{Page-wise matrix transpose}
 This has two flavors, one with homomorphisms and one with tensors:
\begin{equation}
\label{layeredTransposeHom}
\Hom(U,V)\otimes W \rightarrow \Hom(V\dual, U\dual)\otimes W
\end{equation}
\begin{equation}
\label{layeredTransposeTen}
U\otimes V \otimes W\rightarrow V\otimes U\otimes W
\end{equation}

By choosing bases and representing these structures by three-dimensional arrays in \matlab, this is the same as applying the command \verb+pagetranspose+.

\subsubsection*{Page-wise scalar multiplication}
As before, we consider $W$ primarily to be a an indexing space with a fixed basis.
In a space $U\otimes W$, we want to perform scalar multiplication by some number in each copy of $U$ indexed by a basis element in $W$.
The collection of these scalars can be thought of as a vector in $W$, so the scalar multiplication is then really the map
\begin{equation}
\label{layeredScalarMultiplication}
\Id_U\otimes \pi: U\otimes W\otimes W\rightarrow U\otimes W.
\end{equation}
Here $\pi$ is the map from \eqref{eq:DiagonalVector}.
Whether the elements in $U$ are represented as vectors, matrices or higher tensors is immaterial here.

\subsubsection*{Page-wise bilinear form evaluation}
The page-wise version of the map \eqref{eq:bilinearGivesMap} induced from a bilinear form is a map like
\begin{equation}
(U\otimes U)^\vee \otimes W\simeq \Hom (U,U\dual)\otimes W.
\end{equation}
Page-wise evaluation of such a bilinear form means that we consider a map with two vector-matrix multiplications, which we can think of as
\begin{equation*}
    \label{eq:layeredBilin}
    \Hom(\RR,U)\otimes\Hom(U,U\dual)\otimes \Hom(U\dual,\RR)\otimes W\rightarrow \RR\otimes W \simeq W.
\end{equation*}

\subsubsection*{Hadamard or element-wise operations}
For any vector space $U$ we can consider the map
\begin{equation}
U\otimes U\dual \simeq U\dual \otimes U\simeq \Hom(U,\RR)\otimes \Hom(\RR,U)\rightarrow \Hom(U,U).
\end{equation}
If we have chosen a basis for $U$, we can identify $U\dual$ and $U$, and also $\Hom(U,U)$ with $U\otimes U$.
We can then compose the above map with $\pi$ to get a map $U\otimes U\rightarrow U$.
If two vectors are expressed in the basis as $a=(a_1,\dots,a_n)^T$ and $b=(b_1,\dots,b_n)^T$, the image of this pair of vectors is expressed as $(a_1b_1,\dots,a_nb_n)^T$.
The reason that we introduce the decomposition above is that it shows naturally the connection with matrix multiplication:
\begin{equation}
a\otimes b\mapsto a\otimes b^T =\begin{pmatrix}a_1\\ \vdots\\a_n\end{pmatrix}\otimes\begin{pmatrix}b_1&\dots&b_n\end{pmatrix}\mapsto\begin{pmatrix}a_1b_1&\dots& a_1b_n\\\vdots & \ddots & \vdots\\a_nb_1&\dots & a_nb_n\end{pmatrix}
\end{equation}
Finally, the map $\pi$ extracts the vector of diagonal elements.
Again, whether the elements in $U$ are represented as vectors, matrices or higher tensors is immaterial here, and also (after composing with $\pi$) the choice of order of $a$ and $b$ is immaterial.

\subsection{Determinants and inverses}
We could present this theory in the framework of multilinear algebra and alternating forms, but for the applications we have in mind, a more down-to-earth approach will suffice.
The cost is, though, that the linearity properties are less obvious.\\

There are several equivalent ways to define determinants.
The quickest is maybe to define, for a linear transformation in $\Hom_\CC(U,U)$ over $\CC$, the determinant as the product of the eigenvalues (with algebraic multiplicities).
Then we know that if the space is real, the determinant  is a real number, even though it can have complex eigenvalues.
We will think of the determinant as a map
\begin{equation}
\det:\Hom(U,U)\rightarrow \RR.
\end{equation}
With a choice of basis, this can be computed as ordinary for matrices, and therefore can also be though of as a map from e.g. $U\otimes U$ to the real numbers.

\subsubsection*{Page-wise determinants}
If we consider the space $\Hom(U,U)\otimes W$, where $W$ has a chosen basis, we can compute the determinant for each page and think of the determinant as a map
\begin{equation}
\label{layeredDeterminant}
\det{}_W:\Hom(U,U)\otimes W\rightarrow W.
\end{equation}

\subsubsection*{Inverses and page-wise inverses}
The subset
\begin{equation}
GL_U = \det{}^{-1}(\RR\setminus 0) \subset \Hom(U,U)
\end{equation}
is a group under composition, the {\em general linear} group of $U$.
The inverse is the ordinary inverse of a matrix if we have a choice of basis.
In the space $\Hom(U,U)\otimes W$, where $W$ has a chosen basis, we can also compute inverses page by page in the subset where the determinant is non-zero in each coordinate.
This subset is the inverse image $\det{}_W^{-1}(W^\ast)$ of the set $W^\ast\subset W$ of elements which are non-zero in each coordinate under the map \eqref{layeredDeterminant}.
This yields a map (of sets that can be represented as three-dimensional arrays)
\begin{equation}
\label{layeredInverse}
\operatorname{inverse}_W : \det{}_W^{-1}(W^\ast)\rightarrow \det{}_W^{-1}(W^\ast).
\end{equation}
In \matlab, this map is given by the command \verb+pageinv+.

\subsubsection*{Transforming normal vectors}
Consider a normal vector $n$ to a boundary component of a polyhedral domain, and let $v$ be any vector in (or parallel to) that boundary component, i.e. $n\cdot v=0$, or as a matrix product, $n^Tv=0$.
Let $A$ be an invertible matrix transforming the domain to another domain, so that the vector $Av$ lies in the corresponding boundary component of the new domain.
We want to transform $n$ by a matrix $X$ so that $Xn$ is again a normal vector.
So we want:
\begin{align*}
    (Xn)\cdot (Av) = 0 & \Longleftrightarrow\\
    (Xn)^TAv = 0 & \Longleftrightarrow\\
    n^TX^TAv=0 &
\end{align*}
We see that setting 
\begin{equation}
\label{eq:transformNormal}
    X = (A^{-1})^T
\end{equation}
works:
\begin{equation*}
    n^TX^TAv=n^T((A^{-1})^T)^TAv =n^TA^{-1}Av= n^Tv=0
\end{equation*}
To conclude, the normal vector is multiplied by $(A^{-1})^T$ in order to get the new normal vector.
This can of course also be computed page-wise.\\



\subsection{A vectorization library}
\label{subsec:lib}
The vectorized functions of the previous section are implemented in a dedicated library. The multiplication is realized by the following functions:
\begin{verbatim}
  amtam(amx,ama)
  avtam(avx,ama) 
  avtav(ava,avb)
  astam(asx,ama)
\end{verbatim}
where: \textbfn{am} stands for 'array of matrices', \textbfn{av} for array of vectors, \textbfn{as} for array of scalars
and \textbfn{t} for the transpose operator. For instance, the original function \textbfn{amtam(amx,ama)} 
implements the page-wise of product of an array of matrices 
\textbfn{amx} and of the page-wise transpose of an array of matrices \textbfn{ama}. It is overriden by a new function 
\textbfn{pagemtimes(amx,'transpose',ama,'none')}. 
For each of these functions, we have a direct link to the underlying linear algebra operations described earlier.
When making this connections explicit, we will always have chosen bases for the vector spaces considered, so there is now no reason to distinguish between a space and its dual, and transposes can be omitted.\\
As an example, consider \verb+astam(asx,ama)+, which implements page-wise scalar multiplication as in \eqref{layeredScalarMultiplication}.
We can think of the two inputs as scalars and matrices indexed in the same way.
If there are $n$ indices and the matrices have size $k\times m$, this means that in \eqref{layeredScalarMultiplication} we used $W =\RR^n$, $U = \RR^k \otimes \RR^m$, \verb+astam+ is the map
\begin{equation}
\Id_{\RR^k \otimes \RR^m}\otimes\pi:\RR^k \otimes \RR^m\otimes \RR^n\otimes \RR^n\rightarrow\RR^k \otimes \RR^m\otimes \RR^n.
\end{equation}
Here the first three tensor factors on the left correspond to \verb+ama+ and the fourth to \verb+asx+.
All the other functions in the library can be considered in a similar manner.\\
Additionally, multiplications of arrays of vectors or matrices with a single object (matrix of vector) are needed. We have the following functions:
\begin{verbatim}
  amsm(ama,smx)
  amsv(ama,svx)
  smamt(smx,ama)
  svamt(svx,ama)
\end{verbatim}
where: \textbfn{sm} stands for a single matrix and \textbfn{sv} for a single vector. This is implemented as creating copies of single objects and utilizing the functions above.
All these operations follow from similar linear algebra constructions, i.e. by composing \eqref{eq:copy}, \eqref{layeredMultiplication} and \eqref{layeredTransposeTen} for suitable spaces.
Formally, we also need to permute factors before applying these maps, i.e. by (repeatedly) using \eqref{swap}.
Page-wise transpose and page-wise inverse are implemented by functions:
\begin{verbatim}
  aminv(ama)
  amt(ama)
\end{verbatim}
identical to \matlab functions \verb+pageinv+, \verb+pagemtranspose+. See also \eqref{layeredInverse} and \eqref{layeredTransposeTen}, respectively. 
A page-wise determinant, as in \eqref{layeredDeterminant}, is implemented as 
\begin{verbatim}
  amdet(ama)
\end{verbatim}
Finally, a page-wise evaluation of a bilinear form, see \eqref{eq:layeredBilin}, is implemented as
\begin{verbatim}
  avtamav(ava,ama,avb)
\end{verbatim}

\subsubsection*{Historical development}
The original \matlab library was developed in 2003 in \cite{Rahman2003} and later exploited in finite element assemblies \cite{AnjamValdman2015, RahmanValdman2013}.
Some original library functions were optimized due to the implicit  (also called arithmetic or broadcasting) expansion feature in R2016b. 
The additional speedup was achieved due to new page-wise functions in R2020b. 
For the convenience of the user, we provide three versions of libraries containing the above-described functions:
\begin{itemize}
    \item The original library \textbfn{library\_vectorization}.
    \item The updated library \textbfn{library\_vectorization\_implicitExpansion} contains the same functions using the \matlab implicit expansion introduced in R2016b.
    \item The newest library \textbfn{library\_vectorization\_pageOperations} incorporate \matlab functions \textn{pagemtimes}, \textn{pagetranspose} and \textn{pageinv} introduced in R2020b.
\end{itemize}
Application of the latest library to existing codes can shorten evaluation times. 
\begin{table}[H]
\begin{center}
\begin{tabular}{| r | r | r | r | r | }
\hline
&& \multicolumn{3}{c|}{library}\\
\cline{3-5} 
&& original & impexp  & page \\
\hline
level  &  K size  & K [s]  & K [s] & K [s] \\
\hline
0 & 64 & 1.0e-02 & 1.1e-02 & 6.5e-03 \\
1 & 343 & 3.7e-03 & 3.3e-03 & 2.4e-03 \\
2 & 2197 & 1.5e-02 & 1.4e-02 & 9.8e-03 \\
3 & 15625 & 1.4e-01 & 8.8e-02 & 7.3e-02 \\
4 & 117649 & 1.0e+00 & 6.5e-01 & 5.0e-01 \\
5 & 912673 & 1.3e+01 & 1.1e+01 & 7.3e+00 \\
\hline
\end{tabular}
\caption{3D assembly of stiffness matrix $K$ using $P1$ tetrahedral elements. Recomputed from \cite{RahmanValdman2013}.}
\label{ta:timesP1_3D}
\end{center}
\vspace{-0.5cm}
\end{table}
An example that recomputes Table 1 of \cite{RahmanValdman2013} is provided in Table \ref{ta:timesP1_3D}. The utilization of the newest library provides a speedup factor of around 2. Assembly times were obtained on a MacBook Air (M1 processor, 2020) with 16 GB memory running \matlab R2023a.

\section{Prototype codes and performance comparison}

\begin{figure}[H]
\centering
\hspace*{-0.5cm}
\begin{minipage}[c]{0.32\textwidth}
\includegraphics[width=\textwidth]{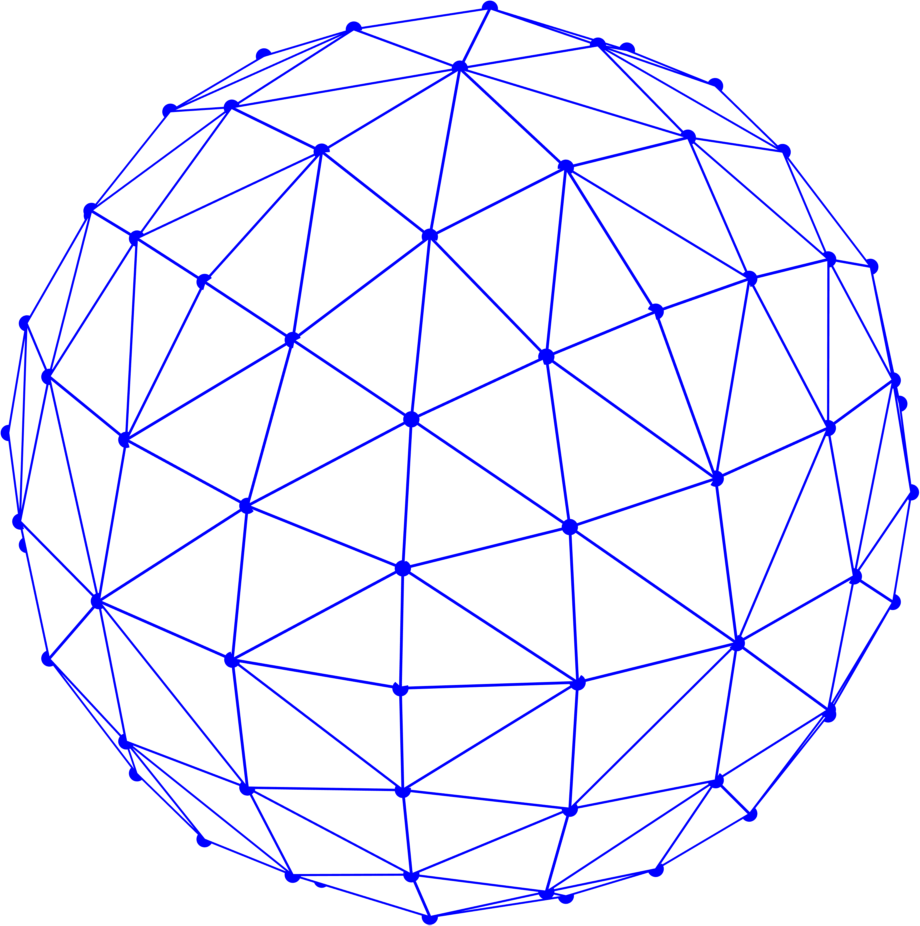}
\end{minipage} 
\hspace*{0.1cm}
\begin{minipage}[c]{0.32\textwidth}
\includegraphics[width=\textwidth]{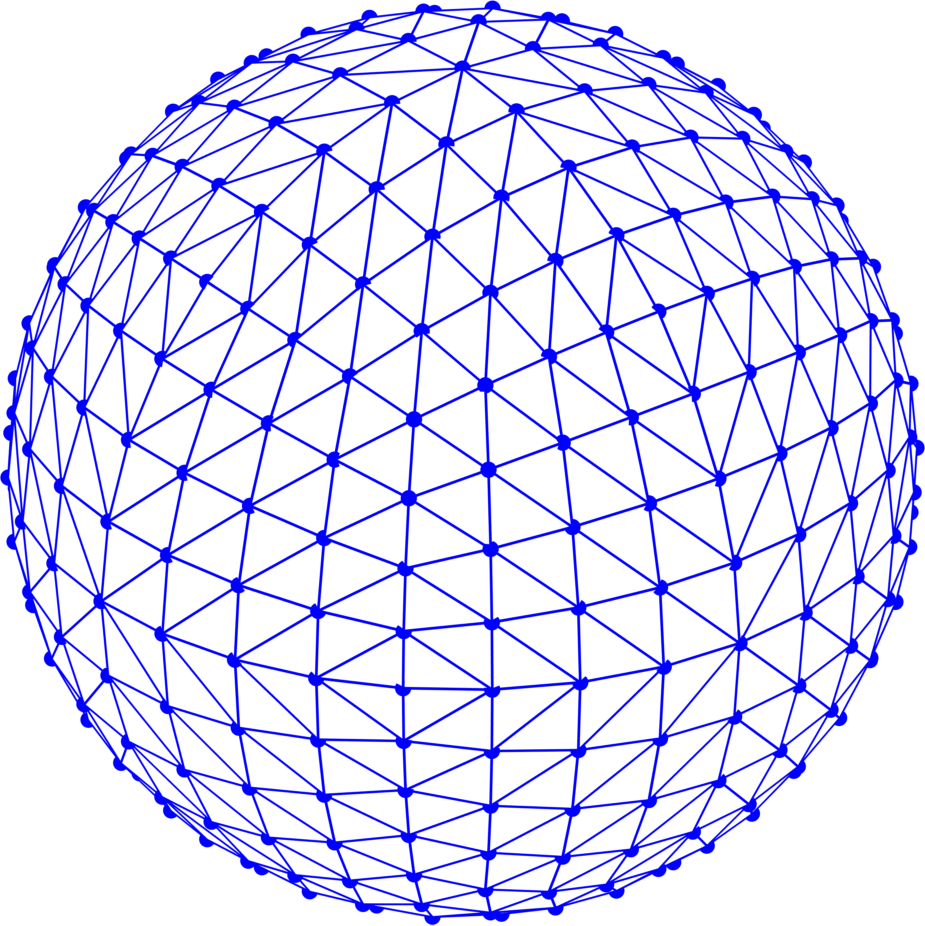}
\end{minipage} 
\hspace*{0.1cm}
\begin{minipage}[c]{0.32\textwidth}
\includegraphics[width=\textwidth]{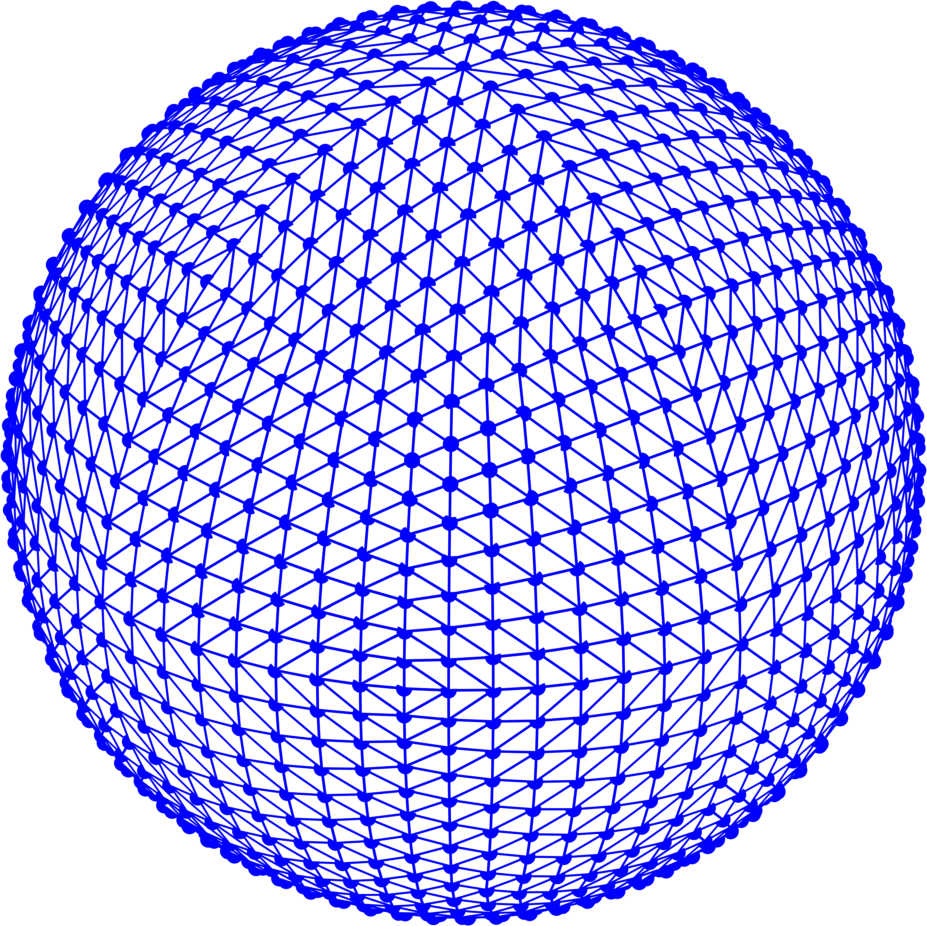}
\end{minipage} 
\caption{Example of 3D uniformly refined tetrahedral meshes (levels 1, 2, 3) of a sphere domain.}
\label{meshes_spheres}
\end{figure}

We assume a three-dimensional ($\dim=3$) domain $\Omega$ approximated by its triangulation $\mathcal{T}$ into closed tetrahedral elements in the sense of Ciarlet \cite{Ciarlet-FEM}. Elements are geometrically specified by their nodes (or vertices) belonging to the set of nodes $\mathcal{N}$. Nodes are also clustered into elements' edges and faces. 
Imported meshes are provided by their own functions or created by a Partial Differential Equation Toolbox of \matlab. 
As an example, we consider a sequence of tetrahedral meshes corresponding to the discretization of the spherical domain $\Omega$ of radius $r=1$, see Figure \ref{meshes_spheres}.
\noindent
Using the script 
\begin{listing}
\begin{lstlisting}[numbers=none]
paramsMesh.level = level;
paramsMesh.r = 1;
[coords,elems] = mesh_sphere(paramsMesh);  
\end{lstlisting}
\end{listing}
for a particular nonnegative integer \textbf{level}, nodes coordinates and tetrahedral elements matrices \textbfn{coords} and \textbfn{elems} are extracted. Numbers of nodes and elements are obtained by 
\begin{listing}
\begin{lstlisting}[numbers=none]
nn = size(coords,1);  % number of nodes 
ne = size(elems,1);   % number of elements
\end{lstlisting}
\end{listing}
 Geometrical properties of spherical meshes are given in the second and third column of Table \ref{tab:volumes}.

\subsection{Structures \coords \, and \vecs} \label{subsec:coords_vecs}
The  first step is an assembly of two 3D matrices. An object
$$ \coords \quad  \text{of size } \; \dim \times \d \times \nt $$
contains the nodes coordinates of every element. Here,  $\dim$ denotes the space dimension (2 or 3), $\d$ is the number of nodes of a single element (3 or 4), and $\nt$ denotes the number of elements. In case of 3D tetrahedra \coords \, is of size $3 \times 4 \times \nt$. Alternatively, it can also be used for storing nodes coordinates of boundary faces which are necessary for the evaluation of the surface integral of a vector field over the domain boundary. In this case,
$$ \coords \quad \text{of size } \; 3 \times 3 \times \nfb \, , $$
where $\nfb$ denotes the number of boundary faces. The second object
$$ \vecs \quad \text{of size } \; \dim \times (\d - 1) \times \nt $$
stores for every element three vectors pointing from the last local node to the first three local nodes. Similarly to \coords, it can also be used for storing vectors of boundary faces which implies this matrix would be of size $3 \times 2 \times \nfb$. \\
Both objects are generated by the function \\
\begin{mm}[caption={The structures \coords \, and \vecs.},label={coords3D}]
function [coords3D,vectors3D] = create_coords3D(coords,elems)
dim = size(coords,2);   % spatial dimension
d = size(elems,2);
ne = size(elems,1);     % number of elements
coords3D = zeros(dim,d,ne);   % 3D matrix of coordinates
for j = 1:d
    coords3D(:,j,:) = coords(elems(:,j),:)';
end
if nargout==2
    vectors3D = coords3D(:,1:end-1,:) - coords3D(:,end,:);
end
\end{mm}

\noindent
\begin{remark}\label{rem:coords3D}
    The assembly construction of \coords \, from the matrices \textbfn{elems} and \textbfn{coords} does not really have a linear algebra interpretation, as the matrix \textbfn{elems} consists of indices.
    On the other hand, the operation producing \vecs \, from \coords \, can be cast in a linear algebra formulation.
    First extract the last coordinate by using the map $\operatorname{last}$ from \eqref{eq:last}:
    \begin{equation}
        \Id\otimes\operatorname{last}\otimes \Id:\RR^{\dim}\otimes \RR^d\otimes \RR^{\nt}\rightarrow \RR^{\dim}\otimes \RR\otimes \RR^{\nt}
    \end{equation}
    Then make copies of this using the map $\operatorname{copy}$ from \eqref{eq:copy}:
    \begin{equation}
        \Id\otimes\operatorname{copy}\otimes \Id:\RR^{\dim}\otimes \RR\otimes \RR^{\nt}\rightarrow \RR^{\dim}\otimes \RR^d\otimes \RR^{\nt}
    \end{equation}    
    If we now compute
    \begin{equation}
        (\Id-\operatorname{copy}\circ \operatorname{last})(coords3D),
    \end{equation}
    we will get an element in $\RR^{\dim}\otimes \RR^d\otimes \RR^{\nt}$ where the last index in the second factor consists only of zeroes (i.e. last coordinate minus last coordinate).
    Finally, we extract everything but this last coordinate by the map (for $d=4$)
    \begin{equation}
        \Id \otimes\begin{pmatrix}
        1&0 &0& 0\\0& 1& 0& 0\\ 0& 0& 1& 0
        \end{pmatrix} \otimes \Id
    \end{equation}
    to retain only the interesting part as an element \vecs \, in $\RR^{\dim}\otimes \RR^{d-1}\otimes \RR^{\nt}$.
\end{remark}

\subsection{Volumes and normals evaluation}\label{subsec:volumes_normals}
\begin{figure}[h]
    \centering
    \includegraphics[width=\textwidth]{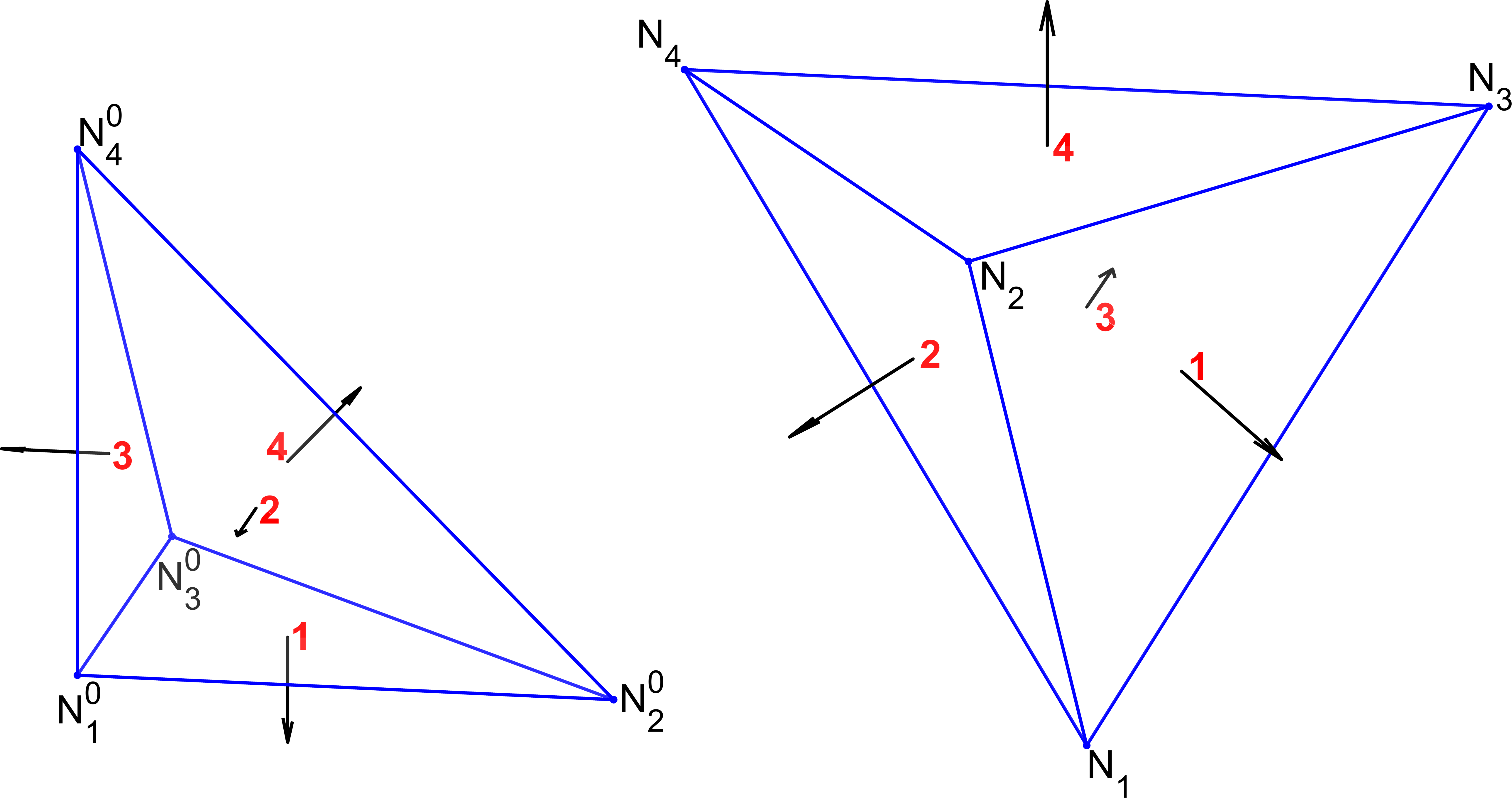}
    \caption{Normals of the reference tetrahedron (left) and normals of a single regular tetrahedron (right).}
    \label{fig:normals_single}
\end{figure}

A reference tetrahedron is defined by four nodes
$$N_1^0 = (0,0,0), \quad N_2^0 = (1,0,0), \quad N_3^0 = (0,1,0), \quad N_4^0 = (0,0,1) $$ and shown in the left part of Figure \ref{fig:normals_single}
It has four faces with outer normals 
that can be stored as columns of a matrix 
\[
\textbf{normalsRef}=
\begin{pmatrix*}[r]
 -1     &    0     &    0  &  1 \\
         0   & -1 &        0  &  1 \\
         0    &     0  & -1 &    1 
\end{pmatrix*}.
\]
\noindent

\begin{example}
For a tetrahedron with nodes
$$ N_1 = \frac{(7, 3, -1)}{4}, \quad N_2 = 
\frac{(7, -2, 4)}{4}, \quad N_3 = \frac{(10, 3, 4)}{4}, \quad N_4 = \frac{(4, 3, 4)}{4}, $$
shown in the right part of Figure \ref{fig:normals_single}, the structure \vecs \, is represented by a matrix (for a single tetrahedron)
\begin{equation}\label{example:one_element}
\frac{1}{4}
\begin{pmatrix*}[r]
 3   &    3     &  6 \\
 0 & -5 &        0   \\
-5    &    0  & 0 
\end{pmatrix*}.
\end{equation}
The determinant of the matrix \eqref{example:one_element} divided by $6$ is equal to 
$-25/64$ and the absolute value of this number corresponds to the tetrahedron volume. To compute normals of the tetrahedron (cf. \eqref{eq:transformNormal}), the formula 
\begin{equation}
\textbf{normals3D}=
\textbf{(vectors3D)}^{-T}\cdot \textbf{normalsRef}   
\end{equation}
is applied  and yields
\[
\textbf{normals3D}=
\begin{pmatrix*}[r]
0 & 0 & -2/3 & 2/3 \\
0 & 4/5 & -2/5 & -2/5 \\
4/5 & 0 & -2/5 & -2/5     
\end{pmatrix*}.
\]
 It should be noted that columns of \textbf{normals3D} are not normalized. 
\end{example}

Given a general tetrahedral mesh and its corresponding structure 
\vecs \,
it is possible to obtain volumes of all tetrahedra at once by a simple script
\begin{listing}
\begin{lstlisting}[numbers=none]
dim=3;                                    % space dimension
meass = amdet(vectors3D)/factorial(dim);  % all volumes 
meas = norm(meass,1);                     % sum of volumes
\end{lstlisting}
\end{listing}
and all normals by 
\vspace{1mm}
\begin{listing}
\begin{lstlisting}[numbers=none]
vectors3D_inv = aminv(vectors3D);              % all inverses
normals3D = amsm(amt(vectors3D_inv),normalsRef); % all normals
\end{lstlisting}
\end{listing}

As an example, Table \ref{tab:volumes} provides times to evaluate the volume of a unit sphere. We observe a quadratic convergence with respect to the mesh size to the exact volume $\frac{4}{3}\pi \approx 4.188790$. Note that the last column "time to evaluate" also includes the times for the mesh generation. The table can be generated by the script
\begin{verbatim}
  benchmark1_volumes_sphere
\end{verbatim}
Consequently, the script
\begin{verbatim}
  benchmark2_normals
\end{verbatim}
evaluates normals for all faces of the same sphere domain for different levels of mesh refinement.
Table \ref{tab:normals} provides the corresponding computational times.

\newcolumntype{d}{>{\hsize=1\hsize}X}
\newcolumntype{s}{>{\hsize=.5\hsize}X}
\newcolumntype{m}{>{\hsize=1.1\hsize}X}
\newcolumntype{Y}{>{\raggedleft\arraybackslash}X}
\newcolumntype{Z}{>{\centering\arraybackslash}X}
\newcolumntype{W}{>{\raggedleft\arraybackslash}m}
\newcolumntype{S}{>{\centering\arraybackslash}s}

\begin{table}[h]
    \centering
    \begin{tabularx}{0.99\textwidth}
    {S |W |W |W |W |W   }
    mesh level & number of elements & number of nodes &   volume & error & time to evaluate \\
     \hline
1 & 384 & 125 & 3.932819 & 2.56e-01 & 1.31e-02 \\
2 & 3072 & 729 & 4.123099 & 6.57e-02 & 3.12e-03 \\
3 & 24576 & 4913 & 4.172259 & 1.65e-02 & 7.22e-03 \\
4 & 196608 & 35937 & 4.184651 & 4.14e-03 & 3.94e-02 \\
5 & 1572864 & 274625 & 4.187755 & 1.04e-03 & 3.32e-01 \\
6 & 12582912 & 2146689 & 4.188531 & 2.59e-04 & 4.42e+00 \\
    \end{tabularx}
    \vspace{-0.1cm}
    \caption{Volume evaluation of a sphere domain.}\label{tab:volumes}
\vspace{0.5cm}
    \centering
    \begin{tabularx}{0.99\textwidth}
       {S |W |W |W |W |W  }
    mesh level & number of elements & number of nodes & number of all faces & number of boundary faces & time to evaluate all faces \\
     \hline
1 & 384 & 125 & 864 & 192 & 1.76e-02 \\
2 & 3072 & 729 & 6528 & 768 & 2.99e-03 \\
3 & 24576 & 4913 & 50688 & 3072 & 7.29e-03 \\
4 & 196608 & 35937 & 399360 & 12288 & 7.67e-02 \\
5 & 1572864 & 274625 & 3170304 & 49152 & 7.35e-01 \\
6 & 12582912 & 2146689 & 25264128 & 196608 & 1.09e+01 \\
    \end{tabularx}
    \vspace{-0.1cm}
    \caption{Normals evaluation of a sphere domain.}\label{tab:normals}
\end{table}

\begin{figure}[h]
\centering
\hspace*{-0.5cm}
\begin{minipage}[c]{0.32\textwidth}
\includegraphics[width=\textwidth]{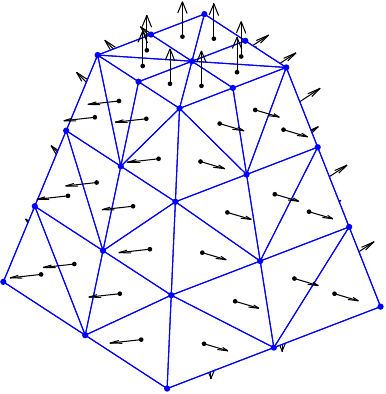}
\end{minipage} 
\hspace*{0.1cm}
\begin{minipage}[c]{0.34\textwidth}
\includegraphics[width=\textwidth]{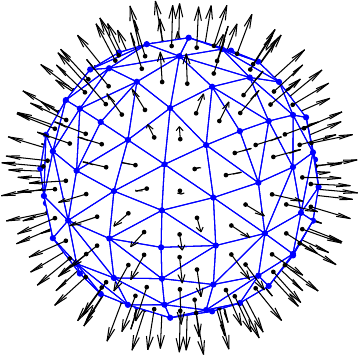}
\end{minipage} 
\hspace*{0.1cm}
\begin{minipage}[c]{0.31\textwidth}
\includegraphics[width=\textwidth]{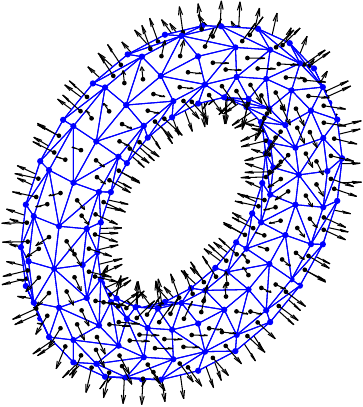}
\end{minipage}
\caption{Examples of 3D meshes and underlying outer normals.}
\label{fig:normals}
\end{figure}

Finally, the script
\begin{verbatim}
  example_normals
\end{verbatim}
generates a 3D mesh of the pyramid, sphere, and torus together with outer normals, whose pictures are shown in Figure \ref{fig:normals}. 
There is an option to compute either normals to all faces (including internal element faces) or only normals corresponding to boundary faces. \\

\noindent

\subsection{Volume integrals} \label{subsec:Gauss}

Provided a domain $\Omega$ and a mass density function $\rho: \Omega \rightarrow \RR$ it is practical to compute an integral
\begin{equation*}
  m=\int_{\Omega} \rho(\x) \, \dxdydz  
\end{equation*}
denoting the total mass of $\Omega$. Given the matrices \textbfn{coords} and \textbfn{elems} for a specific triangulation of $\Omega$ and the mass density function, one can evaluate the total mass of the body using the code below
\vspace{2mm}

\begin{mm}[caption={Gauss quadrature for a volume integral.},label={volume_integral}]
gqo = 3;  % a quadrature order
[ip,w] = Gauss_points(gqo,dim); % barycentric coordinates of quadr. points and weights
coords3D = create_coords3D(coords,elems);
X_ip = amsm(coords3D,ip);  % 3 x nip x ne
rho = @(X) X(1,:,:).^2 + X(2,:,:).^2; % mass density
mass = GI(rho(X_ip),w,volumes); % the total mass
\end{mm}

\noindent
The procedure consists of the following steps:

\begin{itemize}
    \item (line 1) the choice of (Gauss) quadrature order which in general depends on the mass density function;
    \item (line 2) a setup of quadrature points and their weights on a reference element;
    \item (line 3) a construction of a 3D matrix \coords \, introduced in Sec. \ref{subsec:coords_vecs};
    \item (line 4) a construction of a 3D matrix \textbfn{X\_ip} of size $\dim \times \nip \times \nt$ which for the $k$-th element stores the coordinates of all Gauss nodes ($\nip$ denotes their number);
    \item (line 5) definition of a mass density function $\rho(\x) = x_1^2 + x_2^2$;
    \item (line 6) Gauss integration performed by the function \textbfn{GI} explained further in Sec. \ref{subsec:GI}.
\end{itemize}

\begin{figure}
    \centering
    \includegraphics[width=0.7\textwidth]{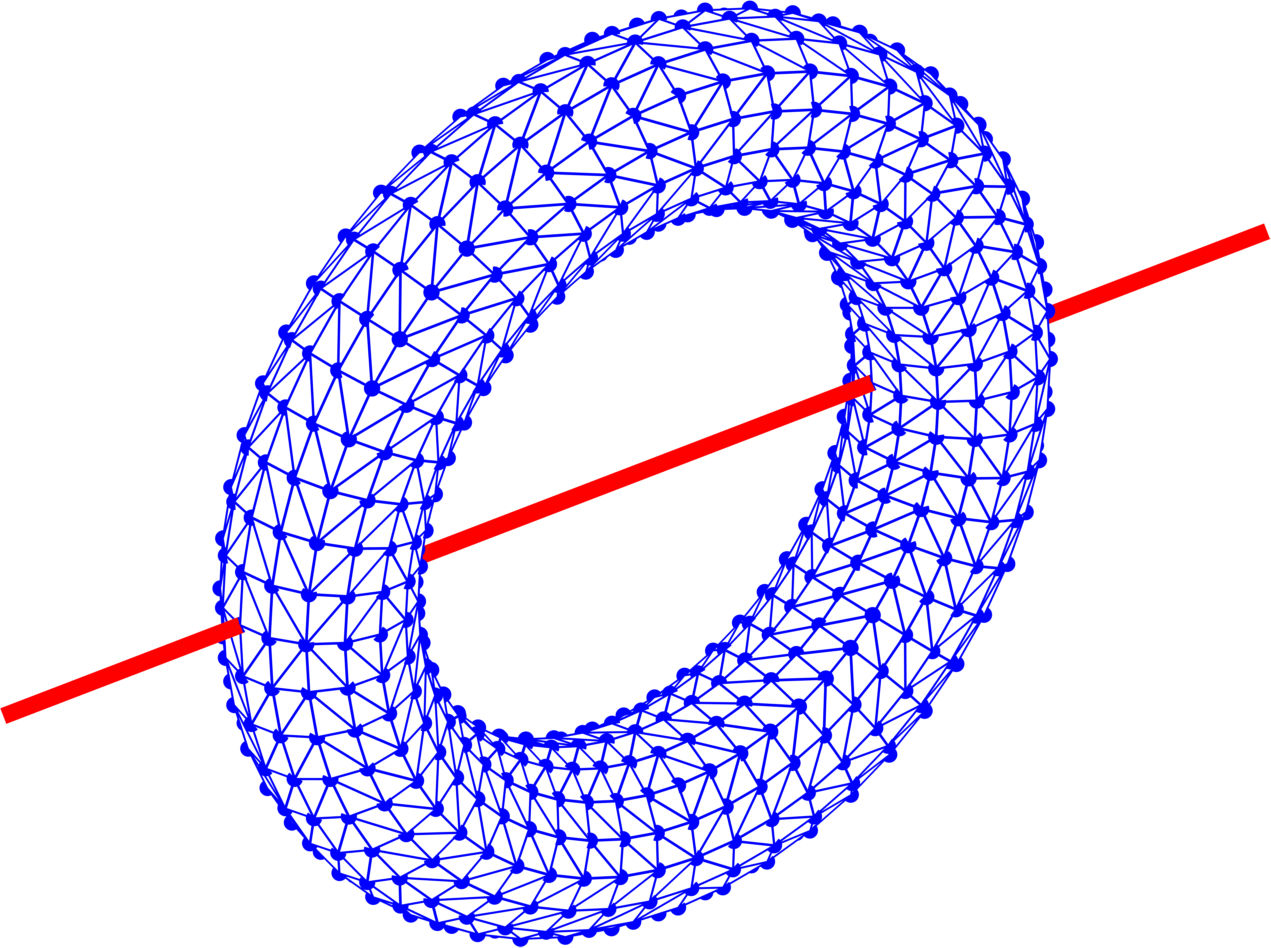}
    \caption{A torus domain rotating around x-axis.}
    \label{fig:torus}
\end{figure}

\begin{remark}
Given the code above, one can also easily evaluate the first and the second moment of the area given by
\begin{eqnarray}
M_i =\int_{\Omega} x_i \, \rho(\x ) \, \dxdydz  , \qquad 
M_{ij} =\int_{\Omega} x_i \, x_j \, \rho(\x) \, \dxdydz .
\end{eqnarray}
It can be done by modifying the last line above (e.g.) with one of the lines below:

\begin{listing}
\begin{lstlisting}[numbers=none]
M1  = GI(rho(X_ip).*X_ip(1,:,:),w,volumes);
M11 = GI(rho(X_ip).*X_ip(1,:,:).^2,w,volumes);
M12 = GI(rho(X_ip).*X_ip(1,:,:).*X_ip(2,:,:),w,volumes);
\end{lstlisting}
\end{listing}
\end{remark}
\noindent
The moments are used for the computation of the mass center coordinates 
$$ x^c_1 = M_1/m, \qquad x^c_2 = M_2/m, \qquad x^c_3 = M_3/m    $$
or the moments of inertia to express the rotational energy of a domain. A benchmark
\begin{verbatim}
  benchmark3_volume_integral
\end{verbatim}
performs several evaluations of the moment of inertia around the x-axis 
\begin{equation}\label{I1}
I_1 = \int_{\Omega} \rho(\x) (x_2^2 + x_3^2) \, \dxdydz = M_{22} + M_{33}    
\end{equation}
for a torus domain shown in Fig. \ref{fig:torus}. Assuming a torus domain given by
\begin{eqnarray*}
 && x_1 = (R + r \cos v)\cos u, \\ 
 && x_2 = r \sin v, \\
 && x_3 = (R + r \cos v)\sin u
\end{eqnarray*}
for parameters $u,v \in (0,2\pi), \, r \in (0, 1/4)$ and $R=1$ and the mass density 
$\rho(x,y,z) = x^2 + y^2$ one can evaluate
$$I_{1} = \frac{2645}{131072}\pi^2 \approx 0.199166.$$ Table \eqref{tab:integration} provides evaluation times for different levels of mesh refinement using the third order of the Gauss quadrature.

\begin{table}
    \centering
    \begin{tabularx}{0.99\textwidth}
      {S |W |W |W |W |W  }
    mesh level & number of elements & number of nodes & value of $I_1$ & error &  time [s] \\
     \hline
0 & 96 & 64 & 0.108588 & 9.06e-02 & 6.11e-03 \\
1 & 576 & 216 & 0.169520 & 2.96e-02 & 1.33e-03 \\
2 & 4992 & 1300 & 0.191582 & 7.58e-03 & 2.68e-03 \\
3 & 38400 & 8100 & 0.197217 & 1.95e-03 & 1.29e-02 \\
4 & 307200 & 57800 & 0.198677 & 4.89e-04 & 5.23e-02 \\
5 & 2482176 & 439956 & 0.199044 & 1.22e-04 & 6.03e-01 \\
6 & 19759104 & 3396900 & 0.199136 & 3.05e-05 & 1.31e+01 \\
    \end{tabularx}
    \vspace{-0.1cm}
    \caption{Evaluation of $I_1$ for a torus domain using volume integration.}\label{tab:integration}
        \vspace{0.5cm}

    \centering
    \begin{tabularx}{0.99\textwidth}
      {S |W |W |W |W |W  }
    mesh level & number of bnd. faces & number of bnd. nodes & value of $I_1$ & error & time [s] \\
     \hline
0 & 128 & 64 & 0.108565 & 9.06e-02 & 1.23e-02 \\
1 & 384 & 192 & 0.169514 & 2.97e-02 & 1.07e-02 \\
2 & 1664 & 832 & 0.191581 & 7.58e-03 & 6.53e-03 \\
3 & 6400 & 3200 & 0.197217 & 1.95e-03 & 6.79e-03 \\
4 & 25600 & 12800 & 0.198677 & 4.89e-04 & 2.10e-02 \\
5 & 103424 & 51712 & 0.199044 & 1.22e-04 & 6.17e-02 \\
6 & 411648 & 205824 & 0.199136 & 3.05e-05 & 2.68e-01 \\
    \end{tabularx}
    \vspace{-0.1cm}
    \caption{Evaluation of $I_1$ for a torus domain using a surface integration.} \label{tab:integration_surface_int}
\end{table}

\subsection{\textbfn{GI} function} \label{subsec:GI}
The key tool of the code above is the \textbfn{GI} function which evaluates an integral of a function or a vector field using Gauss quadrature. It is given by the code below
\vspace{1mm}
\begin{mm}[caption={The \textbfn{GI} function.},label={GI}]
function value = GI(fip,w,sizes,normals)
f_elems = reshape(amsv(fip,w),size(fip,1),size(fip,3))';
if nargin==4
  f_elems = sum(f_elems.*normals,2);  % scalar product 
end
value = sum(sizes.*f_elems);  % the integral value
\end{mm}

\noindent
and has the following inputs:
\begin{itemize}
    \item \textbfn{fip} is a 3D matrix of size $\d \times \nip \times \nt$ storing for every element the values of $f(\x)$ in all Gauss points;
    \item \textbfn{w} is a vector of Gaussian quadrature weights;
    \item \textbfn{sizes} is a vector of elements' sizes (lengths for 1D, areas for 2D, volumes for 3D);
    \item \textbfn{normals} is a matrix of size $\nt \times \dim$ storing outer normals of all elements. It must be provided for the integration of a vector field over a hyperplane.
\end{itemize}
The body of this function consists of the following steps:
\begin{itemize}
    \item (line 2) evaluating a matrix \textbfn{f\_elems} of size $\nt \times \d$ which for every element and every component of $f(\x)$ stores its averaged value over the Gauss points with respect to the corresponding weights;
    \item (line 4) if $f(\x)$ is a vector field, the scalar products with the corresponding normals are calculated and summed over the components of $f(\x)$. In this case the relation $\d = \dim$ holds;
    \item (line 6) the final value of the integral is given by the scalar product of \textbfn{f\_elems} and \textbfn{sizes}. At this stage, \textbfn{f\_elems} is always of size $ne \times 1$.
\end{itemize}
The main advantage of this function lies in the ability to integrate both scalar and vector functions in any spatial dimension.

\subsection{Surface integrals}
Given a domain $\Omega$ with a piecewise smooth boundary $\partial \Omega$ and a vector field
$$F(\x) = \big(F_1(\x),F_2(\x),F_3(\x)\big) \, , \quad \x \in \partial \Omega \, ,$$ one can use boundary normals of Section \ref{subsec:volumes_normals} 
to evaluate
\begin{equation} \label{surf_int}
\int_{\partial \Omega} F(\x) \cdot \overrightarrow{n} \dS \, .
\end{equation}
\begin{figure}[h]
    \centering
    \includegraphics[width=0.7\textwidth]{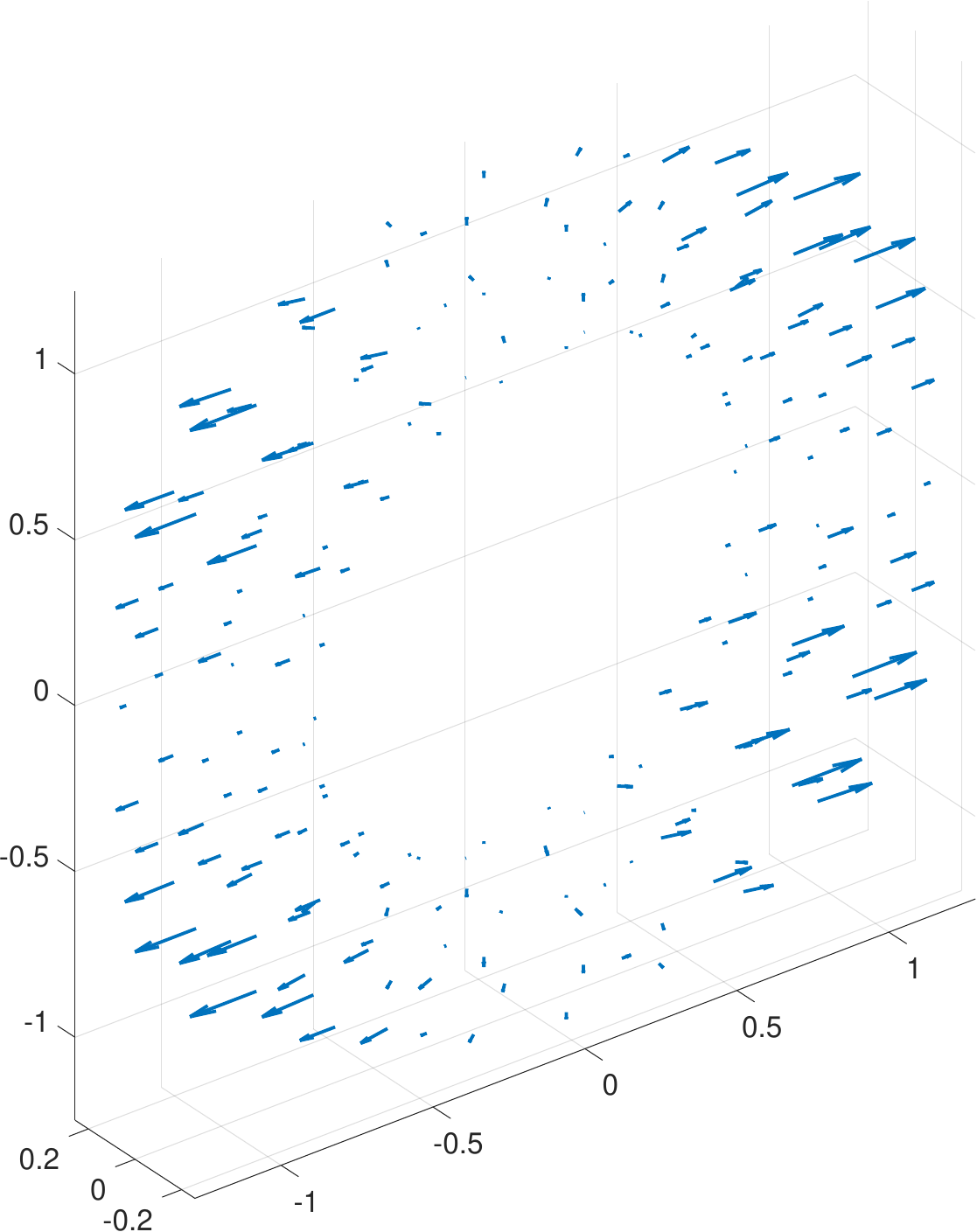}
    \caption{Vector field \eqref{vector_field} depicted on a torus domain.}
    \label{fig:vector_field}
\end{figure}
Since the divergence theorem 
reforms \eqref{surf_int} by
\begin{equation}
\int_{\partial \Omega} F(\x) \cdot \overrightarrow{n} \dS = \int_{\Omega} \nabla \cdot F(\x) \, \dxdydz \, 
\end{equation}
it is possible to recompute all volume integrals of  Sec. \ref{subsec:Gauss} by surface integrals. 
For a vector field (see Fig. \ref{fig:vector_field})
\begin{equation} \label{vector_field}
F(\x) = \Big(\frac{x_1^3 \, (x_2^2 + x_3^2)}{3}, \frac{x_2^5}{5}, \frac{x_2^2 \, x_3^3}{3}\Big)
\end{equation}
one can easily show
$$ \nabla \cdot F(\x) = (x_1^2 + x_2^2) \, (x_2^2 + x_3^2) = \rho(\x) \, (x_2^2 + x_3^2). $$
Therefore the integral \eqref{surf_int} for $F(\x)$ from \eqref{vector_field} corresponds to the moment of inertia $I_1$ of \eqref{I1}. 
The code below evaluates \eqref{surf_int}:
\vspace{2mm}
\begin{mm}[caption={Gauss quadrature for a surface integral.},label={surface_integral}]
gqo = 3;  % a quadrature order
[ip,w] = Gauss_points(gqo,dim-1); % barycentric coordinates of quadr. points and weights
coords3D = create_coords3D(coords,facesB);
X_ip = amsm(coords3D,ip);
F=@(X) [X(1,:,:).^2; X(2,:,:).^2; X(3,:,:).^2]; % vector field
value = GI(F(X_ip),w,areasB,normals);
\end{mm}

\noindent
Here, boundary faces \facesB \, together with their areas \areasB \, are used. The evaluation of $I_1$ over a surface of a torus is done in benchmark
\noindent
\begin{verbatim}
  benchmark4_surface_integral
\end{verbatim}
 Evaluation times are provided in Tab. \ref{tab:integration_surface_int}. Clearly, the numerical values of $I_1$ as well as the corresponding errors are the same as in Tab. \ref{tab:integration}. However, evaluation of the surface integral is significantly faster than the corresponding volume integral.

\subsection{Finite element method}
The finite element method (FEM) is a numerical tool for the solution of partial differential equations \cite{Ciarlet-FEM}. We can easily apply our vectorization concepts to the class of iso-parametric elements. A reference element in local coordinates $\ksi=(\xi_1,\xi_2,\xi_3)$ is transformed to a real finite element in global coordinates $\x = (x_1,x_2,x_3)$  with use of shape
functions $\{\varphi_i\}$.  Then mapping between the global and the reference coordinate systems can be given by the following relation called the iso-parametric property,
\begin{eqnarray}
 x_1 = \sum_i \varphi_i(\ksi) x_{1,i} \, , \quad
 x_2 = \sum_i \varphi_i(\ksi) x_{2,i} \, , \quad 
 x_3 = \sum_i \varphi_i(\ksi) x_{3,i} \, . \nonumber 
\end{eqnarray}
Here, $\x_i = (x_{1,i},x_{2,i},x_{3,i})$ is the i-th point on a general element in the global coordinate system corresponding to the shape function $\varphi_i$. The formula above is defined in three-dimensional space only but can be easily reduced to two-dimensional space as well. Consequently, the same codes are going to work both in two- and three-space dimensions unless stated otherwise.

The shape functions are defined on reference elements. We consider linear (known as P1) and quadratic (known as P2) shape functions $\{ \varphi_i \}$ defined on a reference triangle or a reference tetrahedron. The reference triangle is given by three vertices $\ksi^1=(0,0), \, \ksi^2=(1,0), \, \ksi^3=(0,1)$. For the construction of quadratic shape functions,  additional vertices $\ksi^4, \dots, \ksi^6$ located at the midpoints of the edges are needed. The reference tetrahedron is given by four vertices $\ksi^1=(0,0,0), \, \ksi^2=(1,0,0), \, \ksi^3=(0,1,0), \, \ksi^4=(0,0,1)$ with additional vertices $\ksi^5, \dots, \ksi^{10}$ for the construction of quadratic shape functions. All shape functions $\{ \varphi_i \}$ satisfy the pointwise equality $\varphi_i(\ksi^j)=\delta_i^j,$ where $\delta$ denotes the Kronecker symbol. A function 
\noindent
\begin{verbatim}
 shape = shapefun (point,etype)
\end{verbatim}
evaluates value of all shape functions in any number of points \textbfn{points} in the reference coordinates system. Similarly, a  function 
\noindent
\begin{verbatim}
 dshape = shapeder (point,etype)
\end{verbatim}
evaluates values of their derivatives. An option  \textbfn{etype} specifies whether linear \textbfn{etype='P1'} or quadratic \textbfn{etype='P2'} shape functions are considered. 

\noindent
Derivatives with respect to the global coordinates are easily calculated from the derivatives with respect to the reference coordinates using the Jacobian matrix. It is performed in the following function:
\vspace{2mm}
\begin{mm}[caption={The function \textbfn{phider} evaluating derivatives of shape functions on all elements. Works both in 2D/3D.},label={phider}]
function [dphi,detj,jac] = phider(coords3D,ip,etype)
    
[dim,nlb,ne] = size(coords3D);
nip = size(ip,2);  % number of integration points

dshape = shapeder(ip,etype);  % local derivatives

dphi = zeros(dim,nlb,nip,ne);  % all derivatives
detj = zeros(nip,ne);          % all Jacobians
jac  = zeros(dim,dim,nip,ne);  % all Jacobi matrices

for i = 1:nip
    tjac = smamt(dshape(:,:,i),coords3D);  % dim x dim x ne
    [tjacinv,tjacdet] = aminv(tjac);  % inverses and dets
    dphi(:,:,i,:) = amsm(tjacinv,dshape(:,:,i));
    detj(i,:) = abs(tjacdet);
    jac(:,:,i,:) = tjac;
end
\end{mm}
The input is provided as the array of matrices \textbfn{coords3D}. Additionally, the integration points' values \textbfn{ip} are specified by positions in the local coordinate system  and \textbfn{etype} the type of element considered. The output objects are:
\begin{itemize}
    \item \textbfn{dphi} is a 4D array containing derivatives of all shape functions in all integration points in all elements;
     \item \textbfn{detj} is a matrix storing Jacobians in all integration points on every element;
    \item \textbfn{jac} is a 4D array containing Jacobi matrices in all integration points in all elements.
\end{itemize}

\begin{remark}\label{rem:phider}
    The loop in Code \ref{phider} is only over the relatively low number of integration points, and is therefore not so costly.
    We could in principle, however, also dispose of this loop using the linear algebra setup.
    Here is a brief sketch of how to do this.
    Start from \textbfn{dshape}, considered as an element in $\RR^{dim}\otimes\RR^{nlb}\otimes\RR^{nip}$, and \textbfn{coords3D}, considered as an element in $\RR^{dim}\otimes\RR^{nlb}\otimes\RR^{ne}$. 
    By permuting tensor factors (the map from \eqref{layeredTransposeTen} used repeatedly) and applying page-wise matrix multiplication (the map from \eqref{eq:pagewisemultGeneral} with $W_1=\RR^{nip},\, W_2=\RR^{ne}$), we get the page-wise construction of \textbfn{jac}.
    To compute \textbfn{jacinv} and \textbfn{detj}, we use the page-wise inverse \eqref{layeredInverse} and determinant \eqref{layeredDeterminant} with $W=\RR^{nip}\otimes\RR^{ne}$.
    Finally, \textbfn{dphi} is computed from \textbfn{dshape} and \textbfn{jacinv} by permuting tensor factors, performing page-wise matrix multiplication, and using the diagonal map \eqref{eq:DiagonalVector}, also in a page-wise manner.
    The resulting \textbfn{dphi} is then an element in $\RR^{dim}\otimes\RR^{nlb}\otimes\RR^{nip}\otimes\RR^{ne}$.\\
    Since the loops in the coming code listings are of a similar nature, we will not make the linear algebra connection explicit.    
\end{remark}

We explain how to assemble bilinear forms in the discretization of second-order elliptic problems. Then we typically need to construct a stiffness matrix $K$ and a mass matrix $M$ defined as
\begin{eqnarray}
K_{ij}&=&\int_\Omega c_K(\x) \nabla \Phi_i \cdot \nabla \Phi_j \, \dxdydz, \\
M_{ij}&=&\int_\Omega c_M(\x) \Phi_i \, \Phi_j \, \dxdydz,
\label{eq:K_M}
\end{eqnarray}
where $\nabla$ denotes the gradient operator and scalar coefficient functions $c_K, c_M: \Omega \rightarrow \R $. This 
extends the functionality of \cite{RahmanValdman2013}, where no coefficients
(eg. $c_K = c_M = 1$) were considered. The vectorized implementation is shown below:
\vspace{2mm}

\begin{mm}[caption={Scalar stiffness matrix for P1 elements.},label={stiff_scalar_P1}]
function [K,K3D] = stiffness_matrixP1(elems,coords,coeffs_fun)

dim = size(coords,2);  % problem dimension
ne = size(elems,1);    % number of elements
nn = size(coords,1);   % number of nodes

gqo = 2;  [ip,w,nip] = intquad(gqo,dim);
coeffs = coeffs_in_ip(coords,elems,coeffs_fun,gqo); % nip x ne

coords3D = create_coords3D(coords,elems);  % dim x nlb x ne
[dphi,detj] = phider(coords3D,ip','P1');

nlb = dim +1;  % number of local basic functions

K3D = zeros(nlb,nlb,ne);
for i=1:nip  % loop over all IPs
    dphi3D = squeeze(dphi(:,:,i,:));  % in one IP
    gradTgrad_i = amtam(dphi3D,dphi3D);  % bilinear form without coefficient
    integrand_i = astam(coeffs(i,:),gradTgrad_i); % bilinear form with coefficient
    K3D = K3D + w(i)*astam(detj(i,:),integrand_i);
end

Y3D = reshape(repmat(elems,1,nlb)',nlb,nlb,ne);
X3D = amt(Y3D);
K = sparse(X3D(:),Y3D(:),K3D(:),nn,nn);
\end{mm}
Code \ref{stiff_scalar_P1} works for both triangular and tetrahedral elements. The order of integration \textbfn{gqo} can be specified in line 7. The corresponding integration (Gauss) points \textbfn{ip} and weights \textbfn{w} are then evaluated automatically for a reference element. If a coefficient function $c_K$ provided by the handle \textbfn{coeffs\_fun} is globally constant, element-wise constant, or element-wise linear, it is enough to choose \textbfn{gqo=1}. Then only one integration point is enough
to ensure the exact integration of the stiffness matrix. In other cases,  \textbfn{gqo} should be chosen as 2 or higher. The only loop of the code starting at line 16 runs over the number of integration points \textbfn{nip}. It sums up the local matrices evaluated at each integration point. The local matrices are assembled in all elements at once and stored in a 3D array \textbfn{K3D} which can be extracted as the second function output. 
A straightforward modification of the code provides extensions to P2 elements and to mass matrices.

\begin{mm}[caption={Scalar mass matrix for P2 elements.},label={mass_scalar_P2}]
function [M,M3D] = mass_matrixP2(elems,coords,coeffs_fun)

dim = size(coords,2);  % problem dimension
ne = size(elems,1);    % number of elements
nn = size(coords,1);   % number of all nodes
nnP1 = max(max(elems(:,1:dim+1))); % number of nodes excluding midedges

gqo = 4;  [ip,w,nip] = intquad(gqo,dim);
coeffs = coeffs_in_ip(coords(1:nnP1,:),elems(:,1:dim+1),coeffs_fun,gqo);  % nip x ne

sizes = sizes_of_elements(coords(1:nnP1,:),elems(:,1:dim+1));
detj_abs = sizes*factorial(dim);  % nip x ne

nlb = (dim+1) + nchoosek(dim+1,2);  % number of local basic functions

phiRef = shapefun(ip','P2');
M3D = zeros(nlb,nlb,ne);
for i=1:nip  % loop over all IPs
    phi2D = phiRef(:,i)*phiRef(:,i)';  % in one IP
    phiTphi_i = repmat(phi2D,1,1,ne);  % bilinear form without coefficient
    integrand_i = astam(coeffs(i,:),phiTphi_i); % bilinear form with coefficient
    M3D = M3D + w(i)*astam(detj_abs,integrand_i);
end

Y3D = reshape(repmat(elems,1,nlb)',nlb,nlb,ne);
X3D = amt(Y3D);
M = sparse(X3D(:),Y3D(:),M3D(:),nn,nn);
\end{mm}
Assembly of mass matrix for P2 elements in Code \ref{mass_scalar_P2} has similar structure as the Code \ref{stiff_scalar_P1}. It works for both triangular and tetrahedral elements as well. In this case the minimum \textbfn{gqo=2} is required even for a globally constant, element-wise constant, or element-wise linear coefficient function $c_K$. In other cases, \textbfn{gqo} should be chosen as 4 or higher. The only loop starting at line 18 running over the number of integration points has the same structure as in Code \ref{stiff_scalar_P1}. Here bilinear forms are given by the multiplication of basis functions instead of their gradients. Therefore, \textbfn{phi2D} in line 19 is two-dimensional, while \textbfn{dphi3D} in line 17 of \ref{stiff_scalar_P1} is three-dimensional. Moreover, jacobians denoted by \textbfn{detj} in line 22 are constant on every element, and therefore are one-dimensional.

\begin{figure}[ht]
\centering
\hspace*{-0.5cm}
\begin{minipage}[c]{0.47\textwidth}
\includegraphics[width=\textwidth]{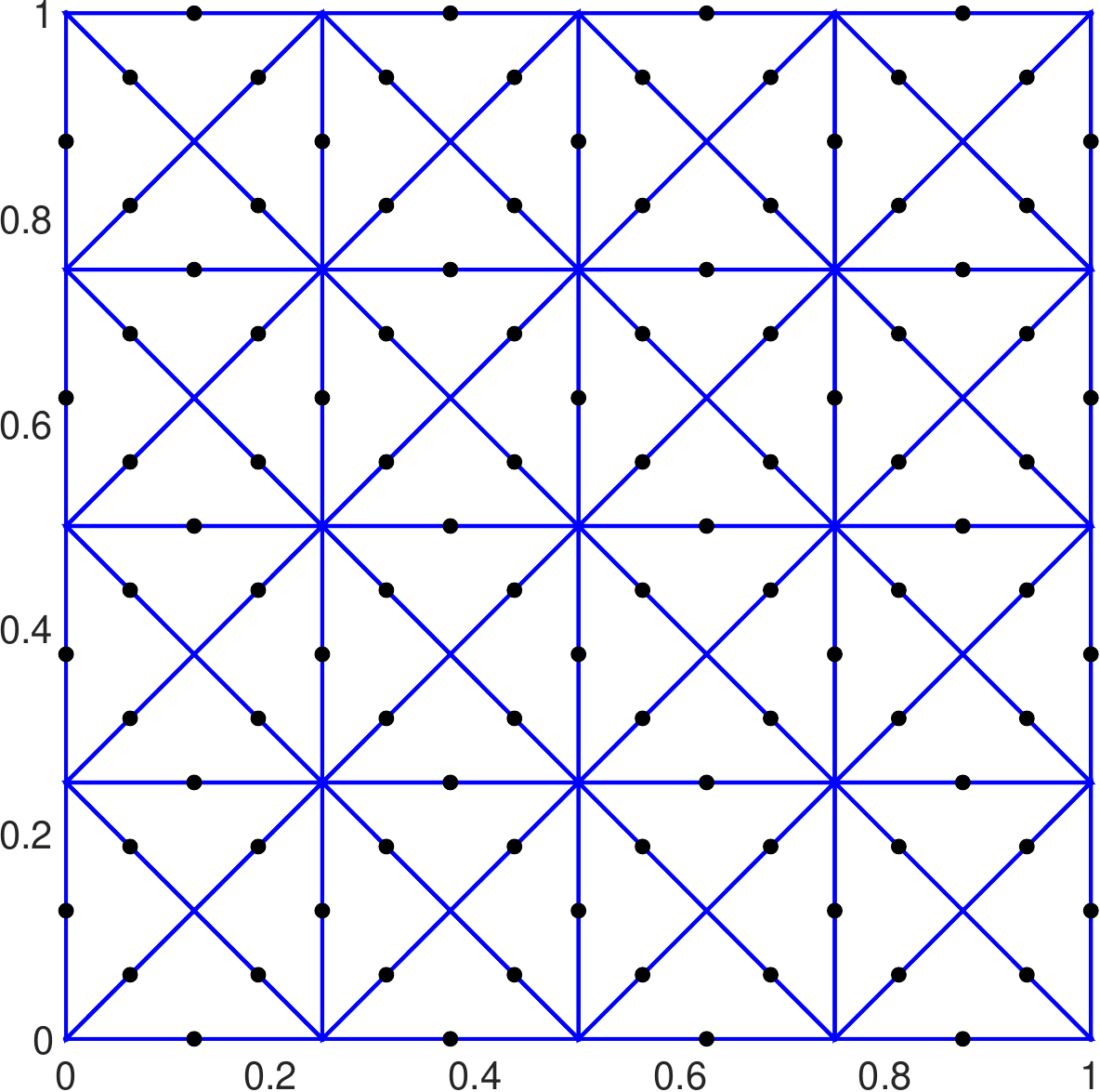}
\end{minipage} \qquad
\begin{minipage}[c]{0.48\textwidth}
\includegraphics[width=\textwidth]{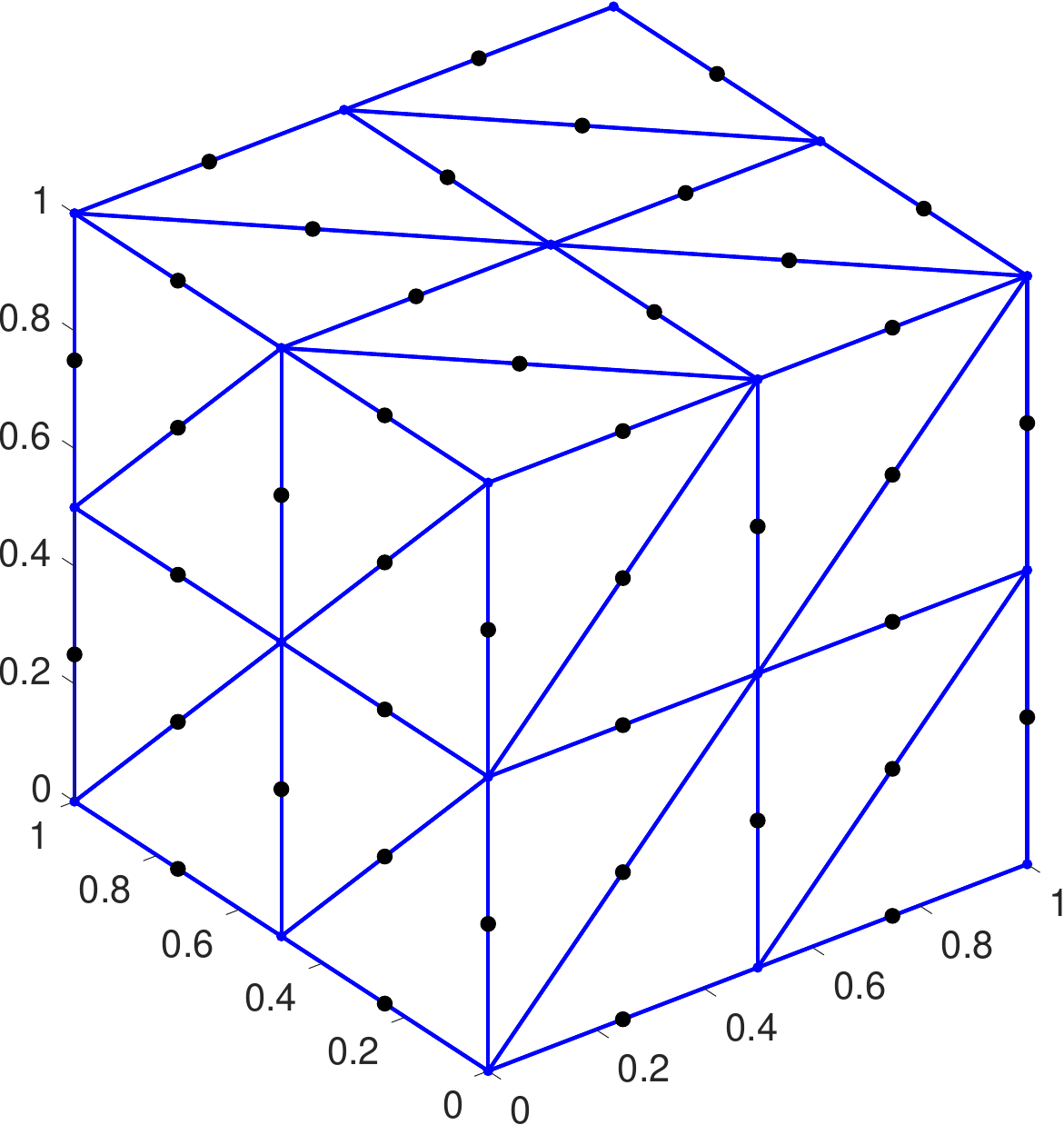}
\end{minipage} 
\hspace*{0.5cm}
\vspace*{-0.3cm}
\caption{Domains and their triangulations. The edges middlepoints (black color) are important in the construction of P2 shape functions.}
\label{fig:meshes_2D_3D}
\end{figure}

\noindent Altogether, we have two pairs of functions:

\noindent
\begin{verbatim}
 [K, K3D] = stiffness_matrixP1(elems,coords,coeffs_fun);
 [M, M3D] = mass_matrixP1(elems,coords,coeffs_fun);

 [K, K3D] = stiffness_matrixP2(elems,coords,coeffs_fun);
 [M, M3D] = mass_matrixP2(elems,coords,coeffs_fun);
\end{verbatim}
The performance of the above assemblies is further discussed for triangulations of a 2D unit square domain  and a 3D unit cube domain shown in Figure \ref{fig:meshes_2D_3D}. 

\noindent
Additionally, we evaluate quadratic forms
\begin{equation}
   \IK := \int_{\Omega}  c_K(\x) \big(\nabla \vx \cdot \nabla \vx\big) \dxdydz \, , \qquad 
   \IM := \int_{\Omega}  c_M(\x) \big(\vx\big)^2 \dxdydz \, 
\end{equation}
for given coefficient functions $\cKx, \cMx$ and a given testing function $\vx$. The above values are approximated by discretized quadratic forms
\begin{equation}
\IK \approx \tilde{v}^T K \tilde{v}, \qquad \IM \approx  \tilde{v}^T M \tilde{v}, 
\end{equation}
where $\tilde{v}$ is a vector of coefficients representing $\vx$ in P1 or P2 basis. The error of such approximation is measured by absolute errors
\begin{equation}
 \eK = \big|\tilde{v}^T K \tilde{v} - \IK\big| \, , \qquad \eM = \big|\tilde{v}^T M \tilde{v} - \IM\big| \, .
\end{equation}

\begin{table}[H]
    \centering
    \begin{tabularx}{0.99\textwidth}
      {S |W |W |W |W |W}
    level & K, M size & $e_K$ & $e_M$ & K [s] & M [s] \\
     \hline
7 & 33025 & 7.30e-04 & 7.05e-05 & 8.64e-02 & 3.96e-02 \\
8 & 131585 & 1.83e-04 & 1.76e-05 & 2.48e-01 & 1.20e-01 \\
9 & 525313 & 4.56e-05 & 4.41e-06 & 9.53e-01 & 5.25e-01 \\
10 & 2099201 & 1.14e-05 & 1.10e-06 & 4.82e+00 & 2.78e+00 \\
11 & 8392705 & 2.85e-06 & 2.75e-07 & 2.98e+01 & 1.46e+01 \\
    \end{tabularx}
    \vspace{-0.1cm}
    \caption{Assembly times of P1 stiffness and mass matrices in 2D.
    } \label{tab:KM_P1_2D}
\vspace{0.5cm}
    \centering
    \begin{tabularx}{0.99\textwidth}
      {S |W |W |W |W |W}
    level & K, M size & $e_K$ & $e_M$ & K [s] & M [s] \\
     \hline
6 & 33025 & 5.87e-08 & 8.49e-09 & 1.04e-01 & 3.17e-02 \\
7 & 131585 & 3.67e-09 & 5.31e-10 & 2.46e-01 & 1.19e-01 \\
8 & 525313 & 2.12e-10 & 3.32e-11 & 1.10e+00 & 5.16e-01 \\
9 & 2099201 & 8.10e-11 & 2.07e-12 & 4.94e+00 & 2.95e+00 \\
10 & 8392705 & 1.07e-10 & 1.28e-13 & 2.96e+01 & 1.28e+01 \\
    \end{tabularx}
    \vspace{-0.1cm}
    \caption{Assembly times of P2 stiffness and mass matrices in 2D.
    } \label{tab:KM_P2_2D}
    \vspace{0.5cm}
    \centering
    \begin{tabularx}{0.99\textwidth}
      {S |W |W |W |W |W}
    level & K, M size & $e_K$ & $e_M$ & K [s] & M [s] \\
     \hline
3 & 729 & 2.70e-01 & 5.04e-02 & 3.22e-02 & 1.49e-02 \\
4 & 4913 & 6.82e-02 & 1.31e-02 & 7.76e-02 & 1.92e-02 \\
5 & 35937 & 1.71e-02 & 3.30e-03 & 3.72e-01 & 1.18e-01 \\
6 & 274625 & 4.28e-03 & 8.28e-04 & 3.99e+00 & 1.88e+00 \\
7 & 2146689 & 1.07e-03 & 2.07e-04 & 6.66e+01 & 2.04e+01 \\
    \end{tabularx}
    \vspace{-0.1cm}
    \caption{Assembly times of P1 stiffness and mass matrices in 3D.} \label{tab:KM_P1_3D}
\vspace{0.5cm}
    \centering
    \begin{tabularx}{0.99\textwidth}
       {S |W |W |W |W |W}
    level & K, M size & $e_K$ & $e_M$ & K [s] & M [s] \\
     \hline
2 & 729 & 6.84e-03 & 5.75e-03 & 6.69e-02 & 1.84e-02 \\
3 & 4913 & 4.00e-04 & 3.88e-04 & 5.86e-02 & 2.18e-02 \\
4 & 35937 & 2.46e-05 & 2.48e-05 & 3.64e-01 & 1.04e-01 \\
5 & 274625 & 1.53e-06 & 1.55e-06 & 4.91e+00 & 2.36e+00 \\
6 & 2146689 & 9.53e-08 & 9.73e-08 & 8.76e+01 & 1.64e+01 \\
    \end{tabularx}
    \vspace{-0.1cm}
    \caption{Assembly times of P2 stiffness and mass matrices in 3D.} \label{tab:KM_P2_3D}
\end{table}

\subsubsection{Assemblies of stiffness and mass matrices in 2D and 3D}
The script
\begin{verbatim}
  benchmark5_assembly_2D
\end{verbatim}
performs assemblies of K and M for nested uniform mesh refinements of the unit square domain and computes approximate values of $\IK$, $\IM$ and their absolute errors $e_K, e_M$. The results are shown in Tables \ref{tab:KM_P1_2D} and \ref{tab:KM_P2_2D}. 
 As a benchmark, we take 
$ \cK(\x) = \cM(\x) = e^{(x_1 + x_2)}, \, \vx = \sin(x_1) \sin(x_2) \,  $
corresponding to the exact values
\begin{equation*}
\begin{split}
   \IK &= \frac{4 \pi^4 \, (e-1)^2 \, (1 + 2\pi^2)}{(1 + 4 \pi^2)^2} 
   \approx 14.5610739535 \, , \\
   \IM &= \frac{4 \pi^4 \, (e-1)^2}{(1 + 4 \pi^2)^2} 
   \approx 0.7021036382 \, .
\end{split}
\end{equation*}
The errors $e_K, e_M$ displayed in the last two columns  decrease quadratically with respect to the mesh size $h$ for P1 elements and with the 4th order for P2 elements. \\

\noindent
The script
\begin{verbatim}
  benchmark6_assembly_3D
\end{verbatim}
performs tests for the unit cube domain.
Here we consider
 $   \cKx = \cMx = e^{(x_1 + x_2 + x_3)} , \,
    \vx = \cos(x_1) \cos(x_2) \cos(x_3) \, $
and can compute that
\begin{equation*}
\begin{split}
   \IK & = \frac{6 \pi^4 \, (e-1)^3 \, (1 + 2\pi^2)^2}{(1 + 4\pi^2)^3} \approx 19.2286024907 \, , \\
   \IM & = \frac{(e-1)^3 \, (1 + 2\pi^2)^3}{(1 + 4\pi^2)^3} \approx 0.6823216700 \, .
\end{split}
\end{equation*}
Tables \ref{tab:KM_P1_3D} and \ref{tab:KM_P2_3D} show assembly times together with the corresponding errors.

\subsubsection{Practical FEM computation} \label{subsec:FEM}

\begin{figure}
\centering
\begin{minipage}[c]{0.43\textwidth}
\includegraphics[width=\textwidth]{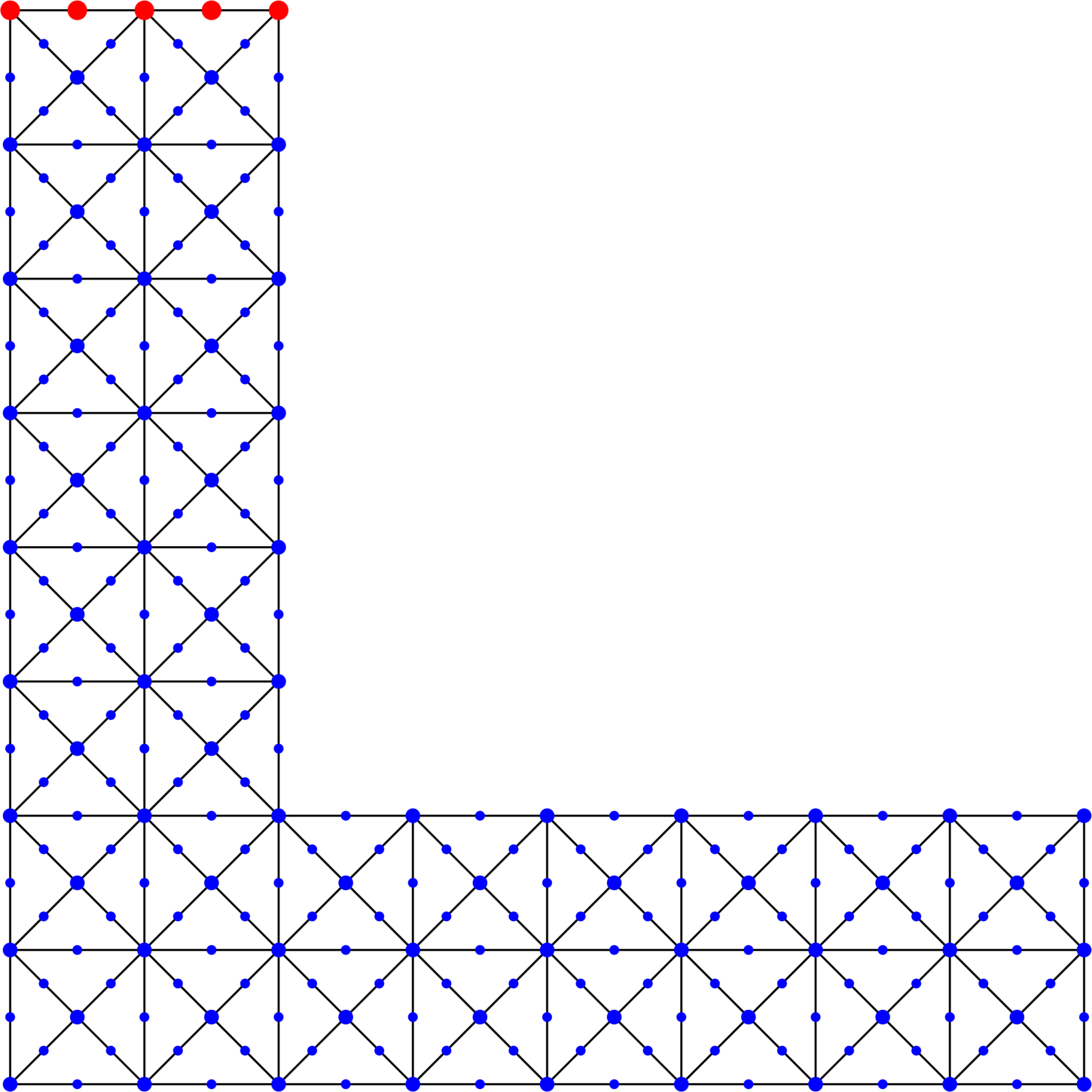}
\end{minipage} \:\:\:
\begin{minipage}[c]{0.52\textwidth}
\includegraphics[width=\textwidth]{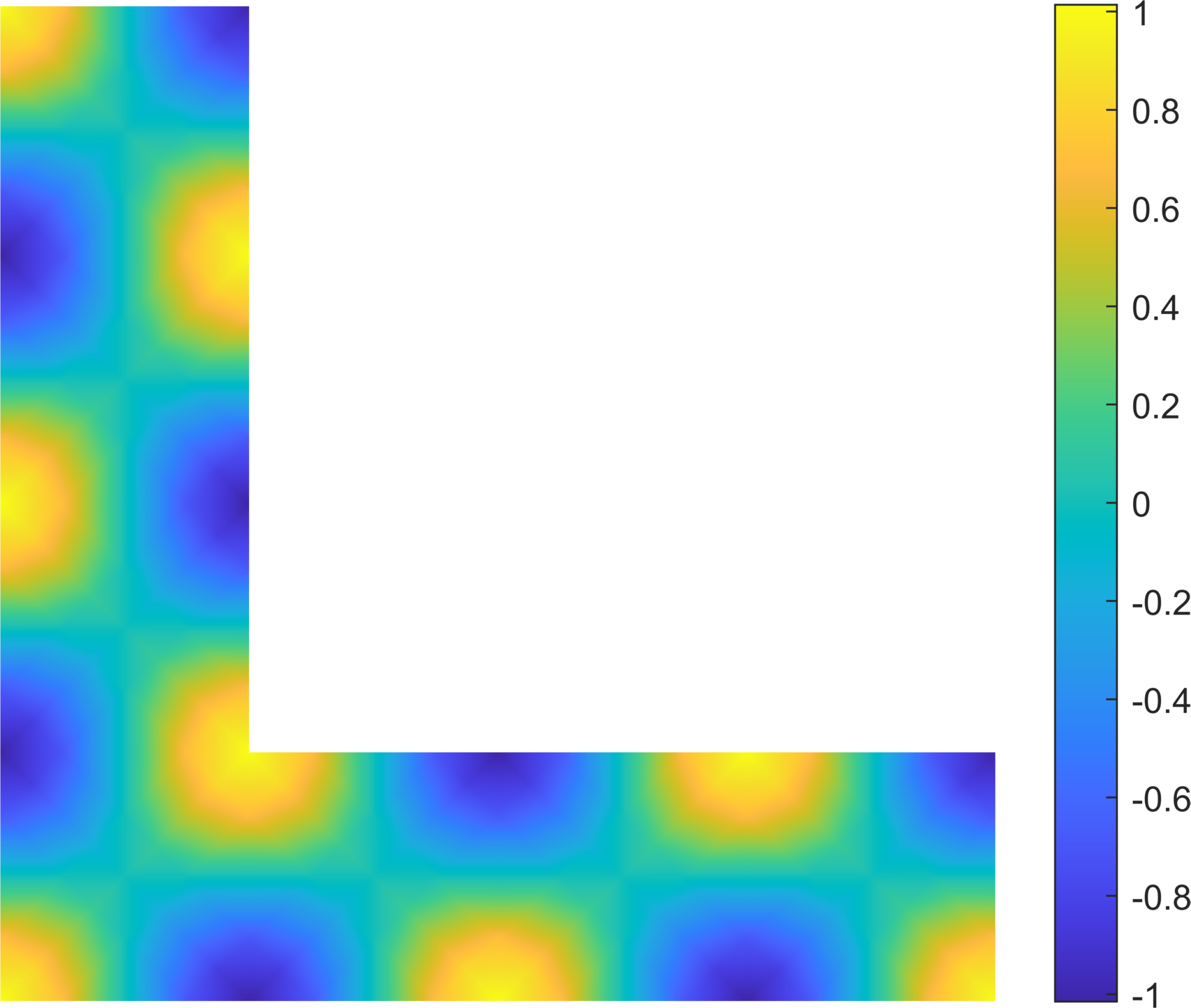}
\end{minipage}
\caption{The computational mesh for P2 elements (left) with $\nt = 112$ and $\nn =257$ (including mid-edge points) and the solution of \eqref{BVP} (right).}
\label{fig:Poisson}
\end{figure}

We focus on solving the full diffusion-reaction boundary value problem
\begin{equation} \label{BVP}
\begin{split}
-\nabla \cdot \big(\cKx \, \nabla \ux\big) + \cMx \, \ux &= \fx \qquad \, \mbox{in} \:\: \Omega \, , \\
\ux &= \uDx \quad \,\,\, \mbox{on} \:\: \Gamma_D \subset \partial \Omega \, , \\
\frac{\partial \uL}{\partial n}(\x) &= 0 \qquad\quad\;\;\, \mbox{on} \:\: \Gamma_N \subset \partial \Omega \, . \\
\end{split}
\end{equation} 
We consider an L-shaped domain $\Omega$ given by the union of rectangles $$(0,\,0.25) \times (0,\,0.25) \, , \;\; (0,\,0.25) \times (0.25,\,1) \, , \;\; (0.25,\,1) \times (0,\,0.25)$$ and shown with its triangulation on the left part of 
Figure \ref{fig:Poisson}.
A non-homogeneous 
Dirichlet boundary condition is assumed on the upper edge $\Gamma_D$ (red nodes) and a zero Neumann boundary condition on the remaining part of the domain boundary $\Gamma_N = \partial \Omega \setminus \Gamma_D$.
One can show that the function
\begin{equation} \label{ux}
    u(\x) = \cos(4 \pi \, x_1) \, \cos(4 \pi \, x_2)
\end{equation}
is the solution of \eqref{BVP} for coefficient functions
$ c_K(\x) = 1 + x_1^2 - x_2 \, , \; c_M(\x) = 1 - x_1 + x_2^2 \, , $ \,
the right-hand side
\begin{equation} 
    \begin{split}
        f(\x) = 8 \pi x_1 \sin(4 \pi x_1) \cos(4 \pi x_2) + \cos(4 \pi x_1) \big( -4 \pi \sin(4 \pi x_2) + \\
        + \big(1 -x_1 + 32 \pi^2 (1 +x_1^2 -x_2) + x_2^2\big) \cos(4 \pi x_2) \big)
    \end{split}
\end{equation}
and the Dirichlet boundary condition
$\uDx = \cos(4 \pi x_1), \; x_1 \in [0,0.25], \; x_2 = 1.$ \\

\noindent
The script
\begin{verbatim}
  benchmark7_BVP_2D
\end{verbatim}
evaluates numerical solutions $\unumx$ (represented by vectors $\uvec$ of coefficients in a finite element basis) of the boundary value problem \eqref{BVP} for different mesh refinements using P1 or P2 finite elements. An example of a P2 solution is shown on the right part of Figure \ref{fig:Poisson}.
Practically, a linear system of equations
\begin{equation}\label{linear_system}
(K + M) \, \uvec = b
\end{equation}
is assembled in which entries of the right-hand size vector $b$ are given by 
\begin{equation} \label{bi}
    b_i = \int_{\Omega} f(\x) \, \Phi_i(\x) \, \dxdydz \, , \qquad i \in \{1, \hdots, \nn\} \, .
\end{equation}
Then it is solved for free entries of $\uvec$, i.e., to those not corresponding to the Dirichlet boundary conditions.

\subsubsection*{Evaluation of the right-hand side vector}
One can evaluate \eqref{bi} by $b = M_0 \, \fvec$, where $M_0$ is the mass matrix corresponding to $\cMx = 1$ and $\fvec$ is a vector of coefficients representing the approximation of $f(\x)$ in a finite element basis. However, this approach is computationally too expensive as long as an additional global mass matrix has to be assembled. Instead, we calculate scalar products of $f(\x)$ and local basis functions ($nlb$ denotes their number)
\begin{equation} \label{fphi}
    \int_{T_k} f(\x) \, \varphi_j(\x) \, , \qquad k \in \{1,\hdots,\nt\} \, , \quad j \in \{1,\hdots,nlb\} \, ,
\end{equation}
on each element. Consequently, any $b_i$ from \eqref{bi} is given as a sum of particular contributions from \eqref{fphi}.

\begin{mm}[caption={The right-hand side vector $b$.},label={rhs}]
function [b,b2D] = rhs_vectorP1(elems,coords,f_fun)

dim = size(coords,2);  % problem dimension
ne = size(elems,1);    % number of elements

gqo = 2;  [ip,w,nip] = intquad(gqo,dim);
coeffs = coeffs_in_ip(coords,elems,f_fun,gqo);  % nip x ne

sizes = sizes_of_elements(coords,elems);
detj_abs = sizes*factorial(dim);  % nip x ne

nlb = size(elems,2);

phiRef = shapefun(ip','P1');
b2D = zeros(nlb,ne);
for i=1:nip  % loop over all IPs
    phi1D = phiRef(:,i);  % in one IP
    integrand_i = coeffs(i,:).*phi1D;
    b2D = b2D + w(i)*detj_abs'.*integrand_i;
end

elems = elems';
b = accumarray(elems(:),b2D(:));
\end{mm}
The code \ref{rhs} evaluates the right-hand side vector $b$ from \eqref{bi} for P1 elements. The only loop starting at line 16 running over the number of integration points evaluates the scalar products \eqref{fphi} that are stored in a matrix \textbfn{b2D}.

\subsubsection*{Evaluation of (local) energies}
\begin{figure}[H]
\centering
\begin{minipage}[c]{0.315\textwidth}
\includegraphics[width=\textwidth]{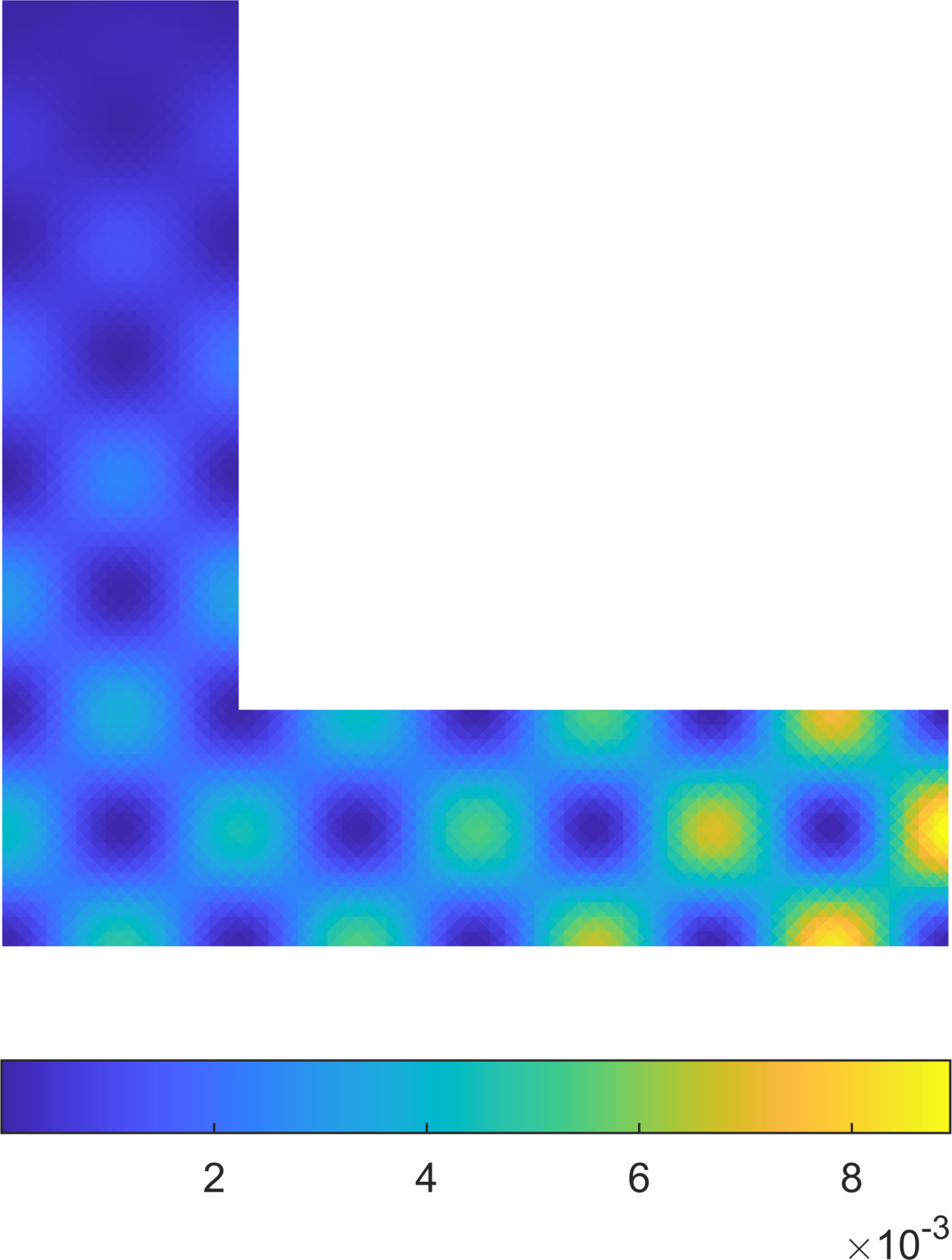}
\end{minipage} \;
\begin{minipage}[c]{0.315\textwidth}
\includegraphics[width=\textwidth]{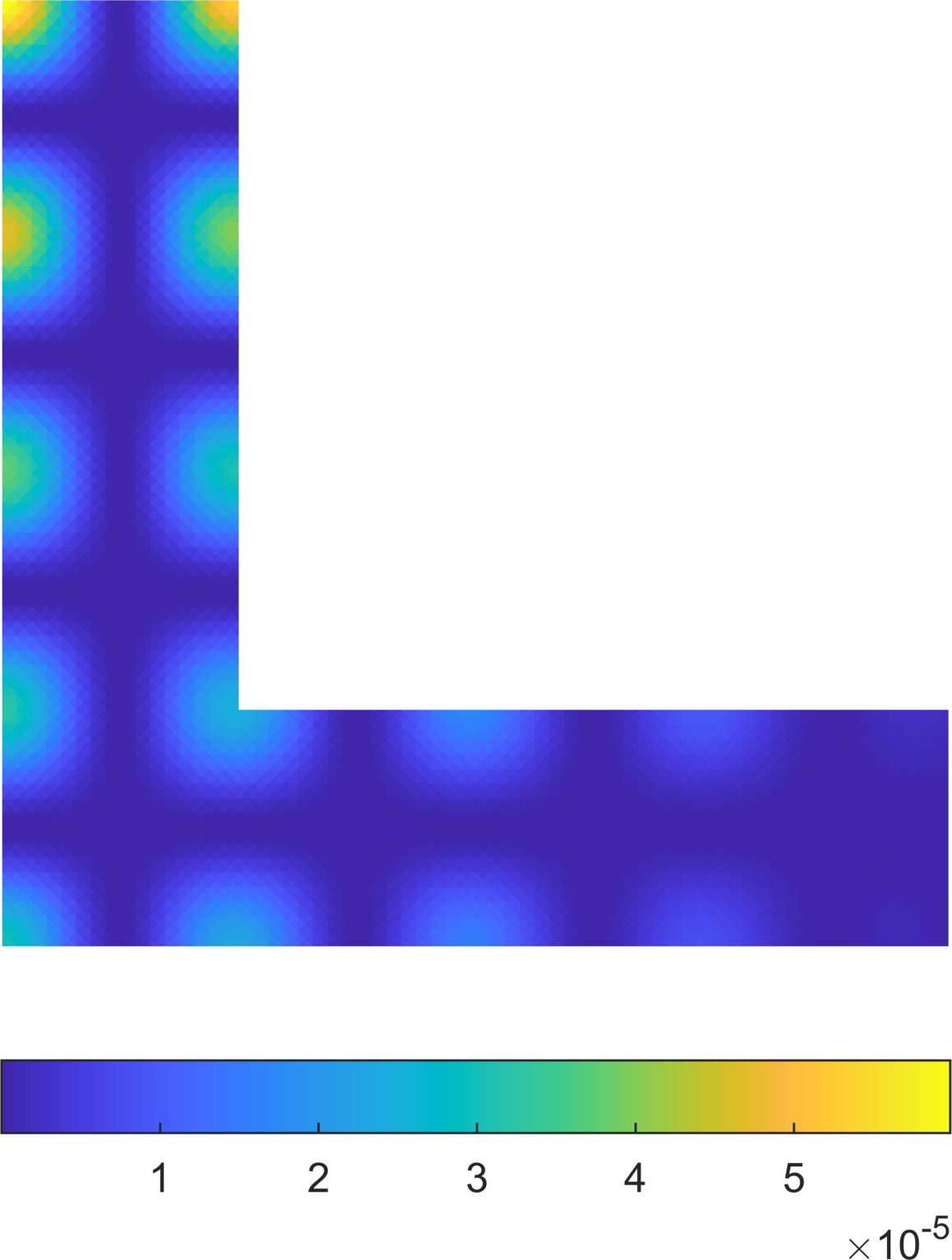}
\end{minipage} \;
\begin{minipage}[c]{0.315\textwidth}
\includegraphics[width=\textwidth]{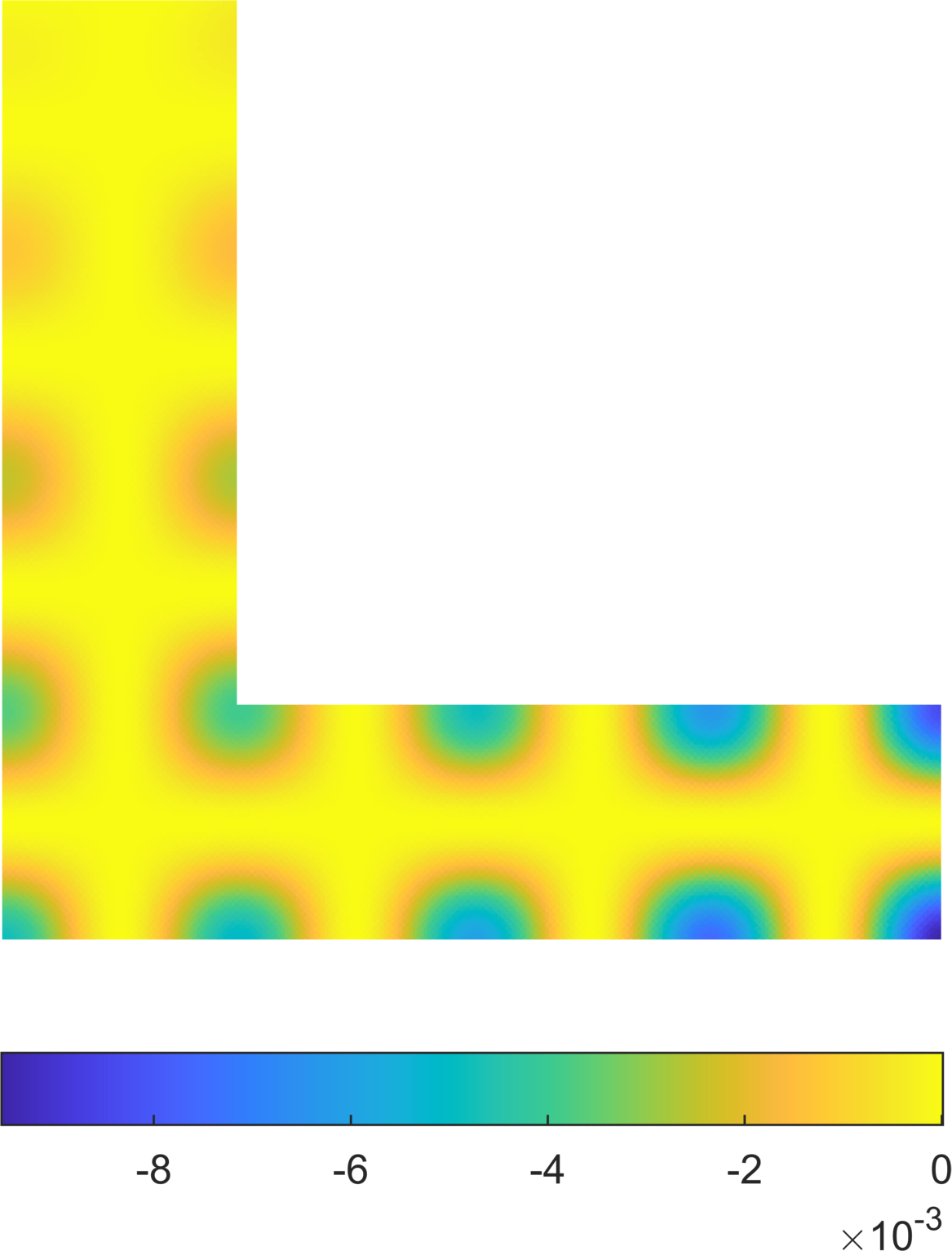}
\end{minipage}
\caption{Approximations of the contributions of $\Ja$, $\Jb$ and $\Jc$ corresponding to the solution \eqref{ux} of the problem \eqref{BVP} with P2 elements and the computational mesh with $7168$ elements.}
\label{fig:Jcomponents}
\end{figure}

An alternative to solve the boundary value problem \eqref{BVP} is to minimize a quadratic energy functional (see \cite{MoVa} for details related to efficient minimization of nonlinear functionals)
\begin{equation}
J(v) = \int_{\Omega} \left( \frac{1}{2} \cKx \|\nabla \vx\|^2 +
\frac{1}{2} \cMx \, \vx^2 \, - \fx \, \vx \right) \dxdydz
\end{equation}
among all testing functions $\vx$ satisfying the Dirichlet boundary condition 
$\vx = \uDx \mbox{ on } \Gamma_D \subset \partial \Omega $.
The minimal value of the energy $J(v)$ is achieved for $v(\x) = u(\x)$, where the exact solution $u(\x)$ is given by \eqref{ux} and 
reads 
 $$ J(u) = \Ja(u) + \Jb(u) + \Jc(u) \approx -14.90302171 \, .$$
Its gradient, reactive and linear parts are
\begin{eqnarray*}
      &&  \Ja(u) = \frac{1}{2} \int_{\Omega} \cKx\ \|\nabla \ux\|^2 \, \dxdydz \,
        = \frac{289 \, \pi^2}{192} \approx 14.85581079 \, , \\
      &&  \Jb(u) = \frac{1}{2} \int_{\Omega} \cMx \, \ux^2 \, \dxdydz \, = \frac{578 \, \pi^2 + 21}{12228 \, \pi^2} \approx 0.04721091674 \, , \\
      &&  \Jc(u) = -\int_{\Omega} \fx \, \vx \, \dxdydz = -\frac{578 \, (32 \, \pi^4 + \pi^2) + 21}{6144 \, \pi^2} \approx -29.80604342 \, .   
\end{eqnarray*}
Using the global matrices $K, M$ and the global vector $b$, values of all three parts are approximated by
\begin{equation} \label{J_comps}
J_1(u) \approx \frac{1}{2} \uvec^T K \uvec \, , \qquad J_2(u) \approx \frac{1}{2} \uvec^T M \uvec \, , \qquad J_3(u) \approx -b^T \uvec \, .
\end{equation}
For any element $T_k$, $k \in \{1, \hdots, \nt\}$, we can also define the local (element-wise) contributions to the energy parts above by formulas
\begin{equation*}
    \begin{split}
        \Jak(u) &= \frac{1}{2} \int_{T_k} \cKx\ \|\nabla \ux\|^2 \, \dxdydz \, , \\
        \Jbk(u) &= \frac{1}{2} \int_{T_k} \cMx \, \ux^2 \, \dxdydz \, , \\
        \Jck(u) &= - \int_{T_k} \fx \, \ux \, \dxdydz \,  
    \end{split}
\end{equation*}
and it holds
\begin{equation*}
    J_i(u) = \sum_{k=1}^{\nt} J_{i,k}(u) \, , \qquad i = 1, \, 2, \, 3.
\end{equation*}
Since our implementation also provides local (element-wise) contributions of $K3D, \, M3D, \, b2D$, it is possible (similarly to \eqref{J_comps} to evaluate the approximations of $\Jak, \, \Jbk, \, \Jck$ for all elements $T_k, \; k \in \{1, \hdots, \nt\}$, at once by
\begin{equation} \label{Jk_comps}
\Jak(u) \, \approx \, \frac{1}{2} \, \uveck^T \, \Kk \, \uveck \, , \quad\; \Jbk(u) \, \approx \, \frac{1}{2} \, \uveck^T \, \Mk \, \uveck \, , \quad\; \Jck(u) \, \approx \, -\bk^T \, \uveck \, .
\end{equation}
Here, $\Kk$ and $\Mk$ are the local stiffness and mass matrices on the $k$-th element, respectively. Similarly, $\bk$ and $\uveck$ are restrictions of $b$ and $\uvec$ on the $k$-th element, respectively. Fig. \ref{fig:Jcomponents} depicts the approximations of energy components \eqref{Jk_comps} for level 4 computational mesh.

\subsubsection{Related projects}
The linear and quadratic functions described above are the simplest of the so-called Lagrange-type shape functions and are the main focus of our paper. Other available implementations that use the vectorization library include:

\begin{itemize}
    \item edge shape functions on triangles or tetrahedra known as Raviart-Thomas and Nedelec elements \cite{AnjamValdman2015},
      \item nodal shape functions on rectangles ensuring the continuity of the first gradient along edges \cite{Valdman2020} known as Bogner–Fox–Schmitt element,
    \item nodal hierarchical-type functions on rectangles \cite{HP} allowing an arbitrary order of a polynomial shape function.
\end{itemize}

Another attempt to avoid the setup of sparse matrices and to work completely with a multidimensional array in terms of iterative solvers is documented in \cite{MarcinkowskiValdman2020}.

\section*{Acknowledgement}
A. Moskovka and J. Valdman were supported by the project grant
23-04766S (GA\v{C}R) on Variational approaches to dynamical problems in continuum mechanics.
\bibliographystyle{abbrv}

\begin{thebibliography}{39}

\bibitem{MATLAB}
The MathWorks Inc.: MATLAB version 9.13.0 (R2022b), Natick, Massachusetts, https://www.mathworks.com .

\bibitem{AlbertyCarstensenFunken1999}  
J. Alberty, C. Carstensen, S. A. Funken: 
{\em Remarks around 50 lines of Matlab: short finite element implementation},
Numer. Algorithms 20(2-3), 117–137, (1999).

\bibitem{AnjamValdman2015}  
I. Anjam, J. Valdman:
{\em Fast MATLAB assembly of FEM matrices in 2D and 3D: edge elements},
Applied Mathematics and Computation 267, 252-263, (2015).

\bibitem{Bader2006}
B. W. Bader, T. G. Kolda:
{\em Algorithm 862: MATLAB tensor classes for fast algorithm prototyping},
ACM Transactions on Mathematical Software, 32(4), 635-653 (2006).

\bibitem{BozorgniaValdman2017}
F. Bozorgnia, J. Valdman:
{\em A FEM approximation of a two-phase obstacle problem and its a posteriori error estimate}, Computers $\&$ Mathematics with Applications 73, No. 3, 419-432, (2017).

\bibitem{Ciarlet-FEM}
P.G.~Ciarlet:
{\em The Finite Element Method for Elliptic Problems},
SIAM, Philadelphia, (2002).

\bibitem{Cuve2016}
F. Cuvelier, C. Japhet, G. Scarella:
{\em An efficient way to assemble finite element matrices in vector languages}, BIT Numer. Math., 56, 833-864 (2016).

\bibitem{FriedrichKruzikValdma2021}
M. Friedrich, M. Kružík, J. Valdman:
{\em Numerical approximation of von Kármán viscoelastic plates},
Discrete and Continuous Dynamical Systems, Series S  14(1): 299-319, (2021).

\bibitem{Koko2007}
J. Koko:
{\em Vectorized Matlab codes for linear two-dimensional elasticity},
Scientific Programming, 15(3), 157-172 (2007).

\bibitem{Bader2009}
T. G. Kolda, B. W. Bader:
{\em Tensor Decompositions and Applications},
SIAM Review, 51(3), 455--500 (2009).

\bibitem{KroemerValdman2019}
S. Krömer, J. Valdman:
{\em Global injectivity in second-gradient Nonlinear Elasticity and its approximation with penalty terms}, 
Mathematics and Mechanics of Solids 24, No. 11, 3644-3673, (2019).

\bibitem{Lit2005}
W. Litvinov, T. Rahman, and R. Hoppe:
{\em Problems of stationary flow of electro-rheological fluids in the cylindrical coordinate system},
SIAM J. Appl. Math., 65(5), 1633--1656 (2005).

\bibitem{Lit2007}
W. Litvinov, T. Rahman, and R. Hoppe:
{\em Model of an electro-rheological shock absorber and coupled problem for partial and ordinary differential equations with variable unknown domain},
Europ. J. Appl. Math., 18, 513-536 (2007).

\bibitem{MacL}
S. MacLane:
{\em Homology. Die Grundlehren der mathematischen Wissenschaften},
Band 114. Springer-Verlag, Berlin-New York, (1963).

\bibitem{MarcinkowskiValdman2020}
L. Marcinkowski, J. Valdman:
{\em MATLAB Implementation of Element-based Solvers},
In: I. Lirkov and S. Margenov (eds): LSSC 2019, LNCS 11958, 601–609, (2020).

\bibitem{MoVa}
A. Moskovka, J. Valdman:
{\em Fast MATLAB evaluation of nonlinear energies using FEM in 2D and 3D: nodal elements}, Applied Mathematics and Computation 424, 127048, (2022).

\bibitem{HP}
A. Moskovka, J. Valdman:
{\em MATLAB implementation of hp finite elements on rectangles using hierarchical basis functions},
PPAM 2022, Lecture Notes in Computer Science (LNCS) 13827, 287-299, (2023).

\bibitem{PaulyValdman2020}
D. Pauly, J. Valdman:
{\em Friedrichs/Poincare Type Constants for Gradient, Rotation, and Divergence: Theory and Numerical Experiments}, 
CAMWA 79,  No. 11,  3027-3067, (2020).

\bibitem{Rahman2003}
T. Rahman:
{\em SERF2D-MatLab (Ver 1.1) - Documentation},
University of Augsburg, (2003).

\bibitem{RahmanValdman2013} 
T. Rahman, J. Valdman: 
{\em Fast MATLAB assembly of FEM  matrices in 2D and 3D: nodal elements}, 
Applied Mathematics and Computation 219, 7151-7158, (2013).

\bibitem{Simo}
D.A. Simovici:
{\em Linear Algebra Tools for Data Mining},
World Scientific, (2012).

\bibitem{Valdman2020}
J. Valdman:
{\em MATLAB Implementation of C1 finite elements: Bogner-Fox-Schmit rectangle},
In: Wyrzykowski R., Deelman E., Dongarra J., Karczewski K. (eds) Parallel Processing and Applied Mathematics. PPAM 2019. Lecture Notes in Computer Science, vol 12044. Springer, Cham, 256-266, (2020).

\end{thebibliography}

\begin{thebibliography}{39}
\bibitem{Carstensen1999}
J. Alberty, C. Carstensen, and S. A. Funken: {\em Remarks around 50 lines of matlab: short finite element implementation}, Numer. Algorithms, 1999, 20, 117-137.

\bibitem{Carstensen2002}
J. Alberty, C. Carstensen, S. A. Funken, and R. Klose: {\em Matlab implementation of the finite element method in elasticity}, Computing, 2002, 69, 236-263.

\bibitem{Koko2007}
J. Koko: {\em Vectorized matlab codes for linear two-dimensional elasticity}, Scientific Programming, 2007, 15(3), 157-172.

\bibitem{Gockenbach2006}
M. S. Gockenbach: {\em Understanding And Implementing the Finite Element Method}, SIAM, 2006.

\bibitem{Persson2004}
P.-O. Persson, and G. Strang: {\em A simple mesh generation in Matlab}, SIAM Rev., 2004, 42, 329-345.

\bibitem{Smith2004}
I. M. Smith and D. V. Griffiths: {\em Programming the Finite Element Method}, 4th ed., John Wiley \& Sons, 2004.

\bibitem{Valdman2009}
J. Valdman:
{\em Minimization of Functional Majorant in A Posteriori Error Analysis
based on H(div) Multigrid-Preconditioned CG Method},
Advances in Numerical Analysis, vol. 2009, Article ID 164519 (2009).

[2] Bahriawati, C., Carstensen, C.: Three matlab implementations of the lowest-order Raviart-Thomas mfem with a posteriori error control. Comput. Meth. Appl. Math. 5,
333–361 (2005)


[5] Funken, S., Praetorius, D., Wissgott, P.: Efficient implementation of adaptive p1-fem in matlab. Comput. Meth. Appl. Math. 11(4), 460–490 (2011)

\end{thebibliography}

\end{document}